\documentclass[aps,prd,showpacs,nofootinbib,floats,floatfix,preprintnumbers,groupedaddress,twocolumn]{revtex4}
\usepackage{bm}
\usepackage{latexsym, environ}
\usepackage{dcolumn, braket}
\usepackage{amsmath,amsfonts,amssymb}
\usepackage{graphicx,epsfig}
\usepackage{fancyhdr}
\usepackage{hyperref, soul}
\usepackage{graphicx,epstopdf}

\def\r{\ref}

\newcommand{\dagg}{\dagger}
\def\s{\sigma} 
\def\o{\omega} 
\def\O{\Omega} 

\def\r{\ref}

\def\a{\alpha}
\def\b{\beta}

\def\g{\gamma}
\def\G{\Gamma}
\def\d{\delta}
\def\D{\Delta}
\def\P{\Phi}
\def\q{\theta}
\def\e{\epsilon}
\def\z{\zeta}
\def\t{\tau}
\def\p{\partial}
\def\f{\frac}

\newcommand{\la}{\langle}
\newcommand{\ra}{\rangle}
\newcommand{\be}{\begin{equation}}
\newcommand{\ee}{\end{equation}}
\newcommand{\bes}{\begin{equation*}}
\newcommand{\ees}{\end{equation*}}

\newcommand{\mc}[1]{\mathcal{#1}}

\newcommand{\bose}{\mathfrak{f}}

\def\l{\left}
\def\r{\right}

\NewEnviron{eqn}{
\begin{align}
\begin{split}
  \BODY
\end{split}
\end{align}
}

\NewEnviron{eqn*}{
\begin{align*}
\begin{split}
  \BODY
\end{split}
\end{align*}
}

\newcommand{\intinf}{\int_{-\infty}^{\infty}} 
\newcommand{\intsinf}{\int_{0}^{\infty}} 

\begin{document}
\title{How robust is the indistinguishability between quantum fluctuation seen from non-inertial frame and real thermal bath}
\author{Chandramouli Chowdhury$^{a,b}$\footnote {\color{blue} chandramouli@alumni.iitg.ac.in,chandramouli.chowdhury@icts.res.in}}
\author{Susmita Das$^a$\footnote {\color{blue} susmita.das@alumni.iitg.ac.in, susmitad210@gmail.com}}
\author{Surojit Dalui$^a$\footnote {\color{blue} suroj176121013@iitg.ac.in}}
\author{Bibhas Ranjan Majhi$^a$\footnote {\color{blue} bibhas.majhi@iitg.ac.in}}

\affiliation{$^a$Department of Physics, Indian Institute of Technology Guwahati, Guwahati 781039, Assam, India\\
$^b$International Centre for Theoretical Sciences, Tata Institute of Fundamental Research, Shivakote, Bengaluru 560089, India}
\date{\today}

\begin{abstract}
We re-advocated the conjecture of indistinguishability between the quantum fluctuation observed from a Rindler frame and a real thermal bath, for the case of a free massless scalar field. To clarify the robustness and how far such is admissible, in this paper, we investigate the issue from two different non-inertial observers' perspective. A detailed analysis is being done to find the observable quantities as measured by two non-inertial observers (one is Rindler and another is uniformly rotating) on the real thermal bath and Rindler frame in Minkowski spacetime.  More precisely, we compare Thermal-Rindler with Rindler-Rindler  and Thermal-rotating with Rindler-rotating situations. In the first model it is observed that although some of the observables are equivalent, all the components of renormalised stress-tensor are not same.  In the later model we again find that this equivalence is not totally guaranteed. Therefore we argue that the indistinguishability between the real thermal bath and the Rindler frame may not be totally true. 
\end{abstract}
\maketitle
\section{Introduction}
General relativity and quantum field theory are the two pillars of modern physics and their coupling led to  great challenges which are yet to yield a fully satisfactory solutions. One of them is the study of quantum field theory in curved space-time which led many researchers to take interest in it. One such topic is the well known Unruh effect which has great importance to understand the Hawking effect \cite{Hawking:1974rv}. In $1976$, Unruh showed that even in flat space-time the particle content of the state of a quantum field is an observer dependent notion \cite{Unruh:1976db}. An observer moving with uniform proper acceleration through Minkowski space-time would see the Minkowski vacuum as a thermal bath of particles, characterised by a temperature $T=a/2\pi$, where $a$ is the proper acceleration of the observer (see \cite{R4} for a review on this topic). 

The origin of the effect in the zero-point fluctuations of the quantum field which are present even in vacuo has been investigated in \cite{R1}. It has been observed that for a detector at rest, the excitations due to zero-point fluctuations are precisely canceled by its spontaneous emission rate and  the Lorentz invariance of the vacuum state ensures that there is still no net excitation for uniform motion. But for the accelerated detector, the correlations in the zero-point fluctuations of the field along the detector's world-line plays the major role in the detector' s response function (we shall see this later) which is no longer being balanced by its own zero-point fluctuations. The detector consequently clicks. The same thing also being investigated by calculating the renormalised stress-tensor of the fields. Since $<T_{ab}>$  vanishes in the Minkowski frame, it must vanish in the Rindler frame as this quantity is a covarint object. This exactly has been shown in literature (see also \cite{R2, R3} for more details).

Apart from understanding this from the accelerated detector's perspective \cite{Padmanabhan:1987rq, Louko:2007mu, Chowdhury:2017ifj}, recently it has been observed that the force due to radiation as measured by the accelerated frame satisfies quantum fluctuation-dissipation theorem \cite{Adhikari:2017gyb}. This observation is quite analogous to the Brownian motion of a particle in a real thermal bath. This led to the conjecture that the vacuum fluctuations seen by a uniformly accelerated observer (Rindler) is equivalent to the thermal fluctuations seen by a static observer in thermal background {\footnote{Similar findings have also been obtained for conformal vacuum, seen by the comoving observers,  in the case of de-Sitter (dS) Friedmann-Lamaître-Robertson-Walker (FLRW) Universe. An extension to ($1+1$) Schwarzschild black hole revels that the particles in Kruskal and Unruh vacuum states, seen by the Schwarzschild static observer, exhibits same fluctuation-dissipation theorem (see \cite{Das:2019aii} for details).}}. A much more deeper analysis of this effect has been investigated. It has been observed that Unruh effect is a microscopic effect rather than a macroscopic one \cite{Buchholz:2014jta}. Moreover, such a scenario is due to the systematic quantum effect induced by the local coupling between the vacuum and the thermometer \cite{Buchholz:2015fqa}.

In order to test such indistinguishability between the quantum fluctuation seen by a non-inertial observer and the thermal fluctuation seen by an inertial observer in thermal bath and how far such is valid, one must study this issue more deeply. {\it In this paper, we try to understand how quantum fields behave in a real thermal bath and whether it can mimic all the phenomena from the perspective of a non-inertial observer by computing different observable quantities}. More precisely, we are interested to examine if another non-inertial observer see the same phenomenon when it looks the real thermal bath and the Rindler frame in Minkowski frame, respectively. Naturally the question arises whether both the situations can produce identical observables with respect to the final same non-inertial frame.     

There have been recent studies in this direction. In \cite{Kolekar:2013aka} it has been shown that the reduced density matrix for a Rindler observer (with acceleration $a$) in thermal bath is symmetric under the interchange of temperature of the bath and $T=a/2\pi$; thereby implies such indistinguishability. Calculation was based on the Unruh modes and has been further elaborated in \cite{Kolekar:2013xua}. Subsequently, the particle number, seen by the Rindler frame in thermal bath, was calculated. Later the same authors demonstrated that the particle number seen by the Rindler-Rindler observer in the Minkowski vacuum is identical to this value \cite{Kolekar:2013hra} with the bath temperature is identified as the temperature perceived by the first Rindler frame. In this analysis they studied both particle number computation by Bogoliubov technique and detector response.
This result shows that one can use Davies-Unruh bath as a proxy for a real thermal bath not only for the initial stage, but also for the next stage where the observer is Rindler one.

Under this circumstances, the most natural question one has to address is to find the robustness of this particular conjecture. Particularly, one needs to investigate if the indistinguishability is still valid with respect to any other non-inertial observer. Moreover as we shall see later that the systems (Rindler frame in Minkowski spacetime and real thermal bath) seen from Rindler frame is not in thermal equilibrium. Therefore, to be more sure of such resemblance, we need to concentrate on more observables. Our aim of this paper is precisely the same.

Here we concentrate on two different models to study this issue. In first example we re-investigate the comparison between the Rindler frame in thermal bath and the Rindler observer in Rindler frame.  We observe that the Green function for the thermal-Rindler case and that for the Rindler-Rindler case are not invariant under time translation and so both of them are not in equilibrium. In that case the thermal flux which is basically the number of particles per unit area is the good quantity for measurement. To measure the thermal flux we adopt the idea of computing the renormalised expectation value of stress tensor of that system in null coordinates. However, we notice that {\it for both the cases the results are not exactly the same}.

Finally, we explore this phenomena with another interesting model introducing rotating frame in a thermal background and shall compare the results with a Rindler-rotating observer. The aim is to find the validity of the indistinguishability with respect to another non-inertial observer other than Rindler. So far we know this has not been investigated in this regard. However, we notice that the Wightman function for the thermal-rotating  observer is time translational invariant, whereas that for the Rindler-rotating observer is not. We find that this has an impact on the observables in these two situations and interestingly, all of them are not exactly identical. The implications are finally discussed.

The organization of the paper is as follows. In the following section; i.e. Section \ref{framework}, we provide the expression of thermal Green's function in position space both in $(1+1)$ and $(1+3)$ spacetime dimensions. A brief description of the detector response is also included in the later part of the same section. These will all provide the main basis of main analysis.  Sections \ref{rind-therm}  and  \ref{Rind-rind} are devoted to calculate different observable quantities for a Rindler observer in real thermal bath and in Rindler-Rindler frame, respectively.  In Section \ref{rot-therm}, we introduce a rotating observer in real thermal bath and find out the detector response function for this case. In the next section, we provide a comparative description of the detector response for a Rindler-rotating observer. Finally in Section \ref{conc}, we summarize our findings and draw a conclusion of our analysis.    

\section{Framework} \label{framework}
In this section, we shall summarise the expressions for thermal Green's function for the free scalar fields in Minkowski spacetime which will be used later for our main purpose. Here the Green's function will be evaluated in coordinate space, both in ($1+1$) and ($1+3$) dimensions. We particularly use Cartesian and Cylindrical coordinates. A short discussion on the Unruh detector response will also be given for the later use of it.    
\subsection{Thermal Green's function in Minkowski frame}
The thermal Wightman function (advanced time) for massless scalar fields in Minkowski spacetime is given by \cite{Mijic:1993wm}:
\begin{eqnarray}
G_{\b}(X_2;X_1) &=&\sum_n\frac{f_n({\bf X_2})f_n^*({\bf X_1})}{2\omega_n} \Big[\frac{e^{i\omega_nT}}{e^{\beta\omega_n}-1}+\frac{e^{-i\omega_n T}}{1-e^{-\beta\omega_n}}\Big]
\nonumber
\\
&\equiv& \sum_n\frac{f_n({\bf X_2})f_n^*({\bf X_1})}{2\omega_n}\Delta_{\beta}(T,\omega_n)~,
\label{Graw}
\end{eqnarray}
where $\beta$ and $\omega_n$ are the inverse temperature of the thermal bath and frequency of the $n^{th}$ mode, respectively. We used the notation $T=T_2-T_1 > 0$ and $X = (T, \mathbf X)$. A derivation of the above expression is presented in Appendix \ref{App4} by considering the thermal scalar fields as collection of canonical ensemble of infinite number of Harmonic oscillators. In the above equation, $f_n({\bf X})$ are the spacial part of the mode solutions of Klein-Gordon equation. Later we shall use this in different situations. Depending upon the case, the relevant expression for $f_n(\bf{X})$ will be substituted and then the sum (for continuum situation, integration) has to be performed. Let us now evaluate this in required forms which are needed for our main analysis.

\subsubsection{$(1+1)$-dimensional spacetime}
As far as we are aware of, in literature ($1+1$), dimensional position space expression for thermal case has not been explicitly mentioned. Therefore, here we shall be little exhaustive to obtain this. Our aim is to find the expression in Cartesian coordinates for massless modes. For that one needs to substitute $f(X)=(1/2\pi)e^{ikX}$ and $\omega=|k| $, where $k$ denotes the wave number. With this, eq.(\ref{Graw}) transforms to 
\begin{eqnarray}
&&G_{\b}(X_2;X_1) = \intinf \f{\,d{k}}{4\pi k} \f{e^{i k \D X}}{e^{\b k} - 1}\l( e^{i k \D T} + e^{\b k} e^{-i k \D T} \r)
\nonumber
\\
&&=\frac{1}{4\pi}\int_{-\infty}^{\infty}\frac{dk}{k(e^{\beta k}-1)}\Big[e^{ik(\Delta T+\Delta X)}+e^{ik(\Delta T - \Delta X)}\Big],
\label{2dg}
\end{eqnarray}
where $\Delta X = X_2-X_1$.
The above integrations are of the following form, 
\be
I_{2D} = \intinf \f{\,d{k}}{k} \f{e^{i k \q}}{e^{\b k} -1}~.
\ee
To evaluate this, we start with the integration below,
\begin{eqn}
I= i \intinf dk \ \f{e^{i k \q}}{e^{\b k} - 1} ~,
\end{eqn}
which after integration with respect to $\theta$ leads to our required result. 
This can be performed by method of complex analysis.
Assuming $\q > 0$, one finds that the upper half of the complex plane is relevant and so the relevant poles which contribute to the integration are at $k = \f{2 \pi n i}{\b}$ with $n$ is positive integers. Then we find
\begin{eqn}
 I = -\frac{2\pi}{\beta}\sum_{n = 1}^{\infty} e^{\frac{- 2\pi n\q}{\b}}=-\f{2\pi}{\b} \f{1}{e^{\f{2\pi \q}{\b}} - 1}~.
\end{eqn}
Integrating this with respect to $\q$ one obtains,
\be
I_{2D}(\q) = -\log\l[ 1 - e^{- \f{2\pi \q}{\b}} \r]~.
\ee
Use of this in (\ref{2dg}) yields the required form: 
\begin{eqn}
G_\beta(X_2;X_1) = - \f{1}{4\pi} \Bigl( &\log\l[ 1- e^{-\f{2\pi}{\b} (\D T - \D X)} \r] \\
&+ \log\l[ 1- e^{-\f{2\pi}{\b} (\D T + \D X)} \r] \Bigr)~.
\label{G2D}
\end{eqn}
In the limit $\beta\rightarrow \infty$ the above reduces to the zero temperature expression
 \be
 \phi(U_2; U_1) \phi(V_2; V_1) = -\f{1}{4\pi}\log\Bigl[(U_2 - U_1)(V_2- V_1) \Bigr]~,
 \label{G01}
 \ee 
where $U=T-X$ and $V=T+X$.
A comment on the $\beta\rightarrow \infty$ which leads to the zero temperature Green's function has been given in Appendix \ref{App3}.

\subsubsection{($1+3$)-dimensional spacetime}
We present thermal Green's function, both in Cartesian as well as Cylindrical coordinates. First let us concentrate on Cartesian case.    

In Cartesian coordinates, $f({\bf X})=(1/2\pi)^3e^{i\bf k\cdot \bf X}$ and so it turns out to be
\begin{equation}
G_{\b}(X_2,X_1)=\int\frac{d^3{\bf k}}{2\omega_{\bf k}} e^{i{\bf{k}} \cdot{\Delta{\bf X}}}\Delta_{\beta}(T,\omega_{\bf k})~,
\label{1}
\end{equation}
where for massless scalar we have $\omega_{\bf k}=|\bf{k}|$. After performing the integration one obtains \cite{Weldon:2000pe}:
\begin{eqn}
&G_{\beta}(X_2,X_1) \\
&= \frac{1}{8\pi\b|\D\mathbf{X}|}\Big[\coth\Big(\frac{\pi}{\beta}(\D T+|\D{\bf X}|)\Big) \\
&\qquad \qquad \qquad \qquad-\coth\Big(\frac{\pi}{\beta}(\D T- |\D{\bf  X}|)\Big)\Big]~,
\label{GCar}
\end{eqn}
where $|{\bf{X}}|=\sqrt{X_1^2+X_2^2+X_3^2}$ with ($X_1,X_2,X_3$) being Cartesian space coordinates and $\D \mathbf{X} \equiv \mathbf{X}_2 - \mathbf{X}_1$. For completeness and clarity, we present a detailed derivation of the above in Appendix \ref{App1}. This also complements the existing calculation, done in \cite{Weldon:2000pe}. For consistency check, if one takes $\b \to \infty$ limit eq.(\ref{GCar}) reduces to the standard zero temperature result, 
\be
G(X_2;X_1) = - \f{1}{(2\pi)^2} \f{1}{(T_2 - T_1)^2 - |{\bf{X}_1} - \mathbf{X}_2|^2}~.
\label{G0}
\ee

  We now turn our attention the derivation in cylindrical coordinates.  From the Klein-Gordon equation in cylindrical coordinates ($\rho,\phi,z$), the normalized mode functions are found out to be,
 \be
 f_n(\mathbf{X}) \equiv f_m(\rho, k_z) = \f{1}{2\pi} J_m(q \rho) \exp(i m \phi + ik _z z)~,
 \label{cyl mode}
 \ee
 with $\o^2 = q^2 + k_z^2$. Here, $m$ denotes the modes along the azimuthal directions (i.e, the conjugate variable of the angular coordinate $\phi$), $q$ denotes the modes functions along $\rho$ direction while $k_z$ is for $z$ direction.  $J_m(q\rho)$ denotes the Bessel function of the first kind of order $m$.
Then (\ref{Graw}) takes the following form:
\begin{eqn}
 &G_\beta(\rho_2,\phi_2,z_2;\rho_1,\phi_1,z_1)\\
 &= \frac{1}{4\pi^2} \sum_{m=-\infty}^{+\infty}\int_0^\infty qdq\int_{-\infty}^{+\infty}\frac{dk_z}{2\omega}
 \ J_m(q\rho_2)J_m(q\rho_1) \\
 &\quad\qquad\qquad\times e^{im\Delta\phi+ik_z\Delta z}
 \left[\frac{e^{i\omega \Delta T}}{e^{\beta\omega}-1}+\frac{e^{-i\omega \Delta T}}{1-e^{-\beta\omega}}\right]~.
\label{new1-1}
 \end{eqn}

We shall use the above results in the subsequent analysis.
\subsection{Detector response: a brief review}
Thermality can be observed in theories by studying the Detector-Response of a Unruh-DeWitt detector. The simplest system of this kind is when one considers a monopole like detector whose motion is described by a classical worldline ($x(\t)$), with $\t$ being the proper time of the detector. The monopole moment of the detector is given by $\mu(\t)$. The detector is assumed to be a two-level system, which makes a transition from some initial energy eigenstate $\ket{E_i}$ to a final energy eigenstate $\ket{E_f}$, when it detects a scalar field. Here we shall assume that the detector is linearly coupled to the scalar field ($\phi(x)$) with the interaction Hamiltonian being,
\be
H_{int} = \mu(\t) \phi[x(\t)]~.
\ee
The time evolution of the detector's moment is governed by its Hamiltonian $H_0$, whose energy eigenstate are $\ket{E_i}$ and $\ket{E_f}$: 
\be
\mu(\t) = e^{i H_0 \t} \mu(0) e^{- i H_0 \t}~.
\ee
The initial and final state of the whole system (detector plus field) is taken as a tensor product of the states of the detector and the field, i.e,
\begin{eqn}
 \ket{I} &= \ket{E_i} \otimes \ket{0}~;\\
 \ket{F} &= \ket{E_f} \otimes \ket{1_p}~.
\end{eqn}
Here, $\ket{0}$ and $\ket{1_p}$ denote vacuum and one particle state of the scalar field with momentum $p$, respectively. Using this information, we can compute the first order transition amplitude of the system from its initial state to final state, 
\be
A(E) = q \intinf\,d{\t} e^{-i E \t} \braket{1_p | \phi[x(\t)] | 0}~.
\ee
Here, $E = E_f - E_i$ and $q = i \braket{E_f|\mu(0)|E_i}$, which only depends on the internal structure of the detector. From this, we can obtain the transition probability by integrating over all possible 1-particle states of the field, 
\be
P(E) = |q|^2 \intinf d{\t_1} d{\t_2}\ e^{-i E (\t_2 - \t_1)} G^+[x(\t_1), x(\t_2)]
\label{power-spectra-defn}
\ee
The positive sign indicates the positive frequency Green's function.

In the case when the {\it{Green's function is time translational invariant}}, we can perform one of the integrals by switching to the coordinates, 
\be
\breve{u} = \t_2 - \t_1, \quad \bar{\t} = \t_2 + \t_1~,
\ee
and divide by, 
\be
|q|^2 T = |q|^2 \intinf \,d{\bar{\t}}~,
\ee
to obtain the {\it response function} of the system, 
\be
R(E) =  \intinf \,d{\breve{u}}\ e^{i E \breve{u}} G^+[\breve{u}]~.
\label{ResponseDefn}
\ee
The above is being used to find the response function in different situations (see \cite{Birrell} for a review on this topic).

It must be emphasised that the applicability of the above expression depends on the translational invarience property of $G^+$; i.e. $G^+$ depends only on the interval ($\breve{u}=\tau_2-\tau_1$) of the detector's proper time. This is usually called as {\it stationary} or {\it equilibrium} system. But if this is not the case, known as {\it non-stationary} or {\it non-equilibrium} system, then it is not possible to use (\ref{ResponseDefn}). In this case a complete analytical analysis of detector's response may not be always possible and consequently any conclusion may not be drawn. Of course, it is aways possible to calculate a finite time Detector response which is given by \cite{Satz:2006kb, Barbado:2012fy}
\begin{eqnarray}
\mathcal{R}(E)=2\int_{-\infty}^{0}d\breve{u}\ Re\left[e^{-i E \breve{u}}\ \mathcal{W}_{R}(\tau_{1},\breve{u})\right]~,   
\label{detc}
\end{eqnarray}
where $\mathcal{W}_{R}(\tau_{1},\breve{u})$ is the regularized Wightman function. This can be used to see the features of the non-stationary situations.
We shall keep this in mind to discuss our main purpose of the present paper.

\section{Rindler observer in thermal bath} \label{rind-therm}
We now consider one of our primary example for understanding quantum effects from a non-intertial frame. The first example that we consider, is a Rindler observer, who is moving through a thermal bath, with an uniform acceleration. 
This model has been studied earlier not only in the perspective of Unruh detector \cite{Costa:1994yx,Kolekar:2013hra}, but also the calculation of particle number, seen from Rindler observer, has been done \cite{Kolekar:2013aka, Kolekar:2013xua}. Consequently it has been argued in \cite{Kolekar:2013xua} that both ways yield the same result. In the second calculation the authors used the Unruh modes to mimic the thermal particles in the Minkowski spacetime while in the earlier one they used the Minkowski modes in determining Green's function. Therefore comparison of these two results may not give the complete story. Keeping this in mind we here calculate everything on the basis of the Minkowski modes which is much more natural to study the present issue {\footnote{A similar ideology has also been mentioned, although not done, at the last paragraph of the paper \cite{Kolekar:2013aka}.}}.

In this section, we revisit this model to deeply investigate different quantities from the perspective of Rindler observer.  For simplicity, the calculations, in this section, are confined in ($1+1$) spacetime dimensions.
The coordinate transformations from Minkowski to Rindler are given by
\begin{eqn}
&T=\frac{e^{a \mathcal{X}}}{a}\sinh(a \mathcal{T})~;\\
&X=\frac{e^{a \mathcal{X}}}{a}\cosh(a \mathcal{T})~,
\label{Rtrans}
\end{eqn}
and under this the Rindler metric takes the form
\begin{eqn}
ds^2&=-d{T}^2+d{X}^2 \\
&= e^{2a \mathcal{X}}(-d\mathcal{T}^2+d\mathcal{X}^2)~.
\label{metric1}
\end{eqn}

\subsection{Particle number}
 The standard methodology for establishing the phenomenon of particle production in quantum field theory, is the evaluation of the {\it number operator} in a particular frame of reference. Since the fields can be represented by infinite collection of Harmonic oscillators (HO), we evaluate this by considering these oscillators immersed in a thermal bath of inverse temperature $\b$, which shall mimic the effects of a free scalar field at finite temperature. Given this similitude, we study the particle spectra (number operator) as seen from an uniformly accelerating observer's (acceleration $= a$) perspective, who is moving through the scalar field (HO) placed in that thermal bath. In the initial computation that we carry out, we shall assume that the system is in thermal equilibrium and use the tools of equilibrium statistical mechanics. The quantity of interest for our calculation then becomes,
\be
\braket{\mathcal{N}}_{\b} = \f{1}{Z} \intsinf \f{\,d{P}}{2 P} \sum_{n = 0}^{\infty} \braket{n| b_P^{\dagg} b_P e^{-\b \mathcal{H}_{\o}} |n}~.
\label{NB} 
\ee
Here, $b_P$ and $b_P^{\dagg}$ are the annihilation and creation operators from the Rindler observer's perspective, respectively.  $\mathcal{H}_{\o} = (a^{\dagg} a) \o$ is the Hamiltonian of the single Harmonic oscillator, with $a$ and $a^{\dagg}$ are its annihilation and creation operators. $Z$ indicates the Partition function given by $Z = \sum_{n}\braket{n| e^{-\b \mathcal{H}_{\o}}|n}$. Evaluation of (\ref{NB}) is straightforward. We present this in Appendix \ref{App2}, which yields 
\be
\braket{N}_{\b} = \f{1}{4} \intsinf \f{\,d{P}}{ P} \Bigl[ 2 \bose_a(P) \bose_{\b}(\o) + \bose_a(P) + \bose_{\b}(\o) \Bigr]~,
\label{NTHR}  
\ee
where $\bose_{\b}(k) = (e^{\b k} - 1)^{-1}$ and $\bose_a(k) = (e^{2\pi k/a } - 1)^{-1}$ denote the Bose-Einstein factors. The same result was obtained in \cite{Kolekar:2013xua} by constructing the thermal density matrix which is being found by integrating out the left modes. There the authors have used the Unruh modes to evaluate the thermal density matrix. On the contrary here we have used Minkowski modes and hence the Minkowski Hamiltonian to evaluate the particle number perceived from the accelerating observer. Although our present procedure is similar in idea with the existing one, but the quantity is slightly different as it has been constructed differently. The significance of appearance of each term is discussed in \cite{Kolekar:2013xua}.

Let us now make some comment on this way of evaluating the particle number. Although this is a simple and interesting result, an issue is there as one used the techniques of equilibrium statistical mechanics. To see this, we shall use the Rindler-Null coordinates, 
\begin{eqn}
\mathcal{U} = \mathcal{T} - \mathcal{X}, \quad \mathcal{V} = \mathcal{T}  +\mathcal{X}
\label{Rindler-Null}
\end{eqn}
In this coordinate system, the metric transforms to, 
\be
ds^2 = - e^{2 a (\mathcal{V} - \mathcal{U}) } d\mathcal{U}d\mathcal{V}~.
\ee
Interestingly, in these coordinates the Green's function decomposes into two parts: one corresponds to outgoing mode and other one is for ingoing modes. The explicit expression is  
\begin{eqn}
&G_{\beta}(\mc{U}_2, \mc{V}_2; \mc{U}_1, \mc{V}_1) =   \braket{\phi(\mc{U}_2) \phi(\mc{U}_1)}_{\b} + \braket{\phi(\mc{V}_2) \phi(\mc{V}_1)}_{\b}
\end{eqn}
where, 
\begin{eqn}
\braket{\phi(\mc{U}_2) \phi(\mc{U}_1)}_{\b} &= -\f{1}{4\pi} \log\Bigl[  1 - e^{\f{2\pi}{a\b} \big( e^{-a \mc{U}_2} - e^{- a \mc{U}_1} \big)}\Bigr]~; \\
\braket{\phi(\mc{V}_2) \phi(\mc{V}_1)}_{\b} &= -\f{1}{4\pi} \log\Bigl[  1 - e^{-\f{2\pi}{a\b} \big( e^{a \mc{V}_2} - e^{a \mc{V}_1} \big)}\Bigr]~.
\label{Wight-TR}
\end{eqn}
Note that the above {\it {is not time translational invariant}} in proper frame of observer
and hence from the accelerated observer's frame the system is not in thermal equilibrium. Therefore a question arises on the viability of the above obtained result.

\subsection{Components of renormalised energy-momentum tensor}
We observed that there exists a problem in the evaluation of number of particles and therefore it is not a good quantity to use for our later purpose. In this situation a better idea for the number of quanta emitted, can be obtained from the thermal {\it flux} as perceived by the accelerating observer. The flux is effectively the number of particles emitted per unit area. The measure of this is best understood, when we study the expectation value of the {\it stress tensor} of the system in the null coordinates, given by eq.(\ref{Rindler-Null}). Using the value of Green's function in these coordinates (see, eq.(\ref{Wight-TR})) and the covariant expression of the energy-momentum tensor for massless scalar fields in ($1+1$) dimensions, 
\begin{equation}
T_{ab}=\nabla_a\phi\nabla_b\phi-\frac{1}{2}g_{ab}\nabla_c\phi\nabla^c\phi~,
\end{equation}
we have a general formula for the renormalised energy-momentum tensor in a background described by the metric,
 \be
 ds^2 = C(\mc U, \mc V) d\mc U d \mc V~.
 \ee
 It is expressed in terms of the contribution arising from the normal ordered value of the stress tensor and the contributions arising from the parallel transport along the direction of point splitting (which is necessary to maintain the expectation values of the stress tensor covariant\footnote{See Chapter 6 of \cite{Birrell} for details. Typically for curved spacetime this expression also has a factor containing the Ricci Scalar, but since we are dealing with the case of flat spacetime, that term naturally vanishes.}.), 
 \be
 \braket{T_{a}^{\ b}[g_{cd}(x)]}_{ren} = \sqrt{-g} \braket{T_a^{\ b}[\eta_{cd}(x)] } + \q_a^b~.
 \label{ST-general:2D}
 \ee
 Here $\braket{T_a^{\ b}[\eta_{cd}(x)] } $ is the renormalised stress tensor corresponding to flat spacetime $\eta_{ab}$, can obtained by performing a simple point splitting  as described in Appendix \ref{App5} and $\q_{ab}$ is the term which ensures covariance. The components of this term are given by 
\begin{eqn}
 \q_{\mc U \mc U} &= - \left( \frac{1}{12\pi} \right) C^{1/2} \p_{\mc U}^2 C^{-1/2}~, \\
 \q_{\mc V \mc V} &= - \left( \frac{1}{12\pi} \right) C^{1/2} \p_{\mc V}^2 C^{-1/2}~,\\
 \q_{\mc U \mc V} &= \q_{\mc V \mc U} = 0~.
 \label{PRD1}
\end{eqn}
For the present case $C$ is given by $C=-e^{2a(\mc V - \mc U)}$. Consequently, we have
\begin{eqn}
    \q_{\mc U \mc U} = - \frac{a^2}{48\pi}, \quad \q_{\mc V \mc V} = - \frac{a^2}{48\pi}~.
   \end{eqn}
Since the first term on the right hand side of (\ref{ST-general:2D}) arises from the normal ordered expansions, for the ease of notations, we shall describe it with $\braket{:T_a^{\ b}(x) :} $~. Evaluating the same for this case (where we should keep in mind that the expectation values are to be taken in the thermal state) we get,  
\begin{eqnarray}
\braket{T_{\mc{UU}}(\mc{U})}_{\b} &&\overset{Ren}{=} \braket{:T_{\mc{UU}}(\mc{U}):}_{\b} + \q_{\mathcal U \mathcal U} \nonumber\\
&&= \f{a^2}{48 \pi} + \f{\pi}{12\b^2} e^{- 2 a \mc{U}} +  \q_{\mathcal U \mathcal U}= \f{\pi}{12\b^2} e^{- 2 a \mc{U}}~; \nonumber \\
\nonumber
\\
\braket{T_{\mc{VV}}(\mc{V})}_{\b} &&\overset{Ren}{=} \braket{:T_{\mc{VV}}(\mc{V}):}_{\b} + \q_{\mathcal V \mathcal V} \label{Tuu-TR} \\
&&= \f{a^2}{48 \pi} + \f{\pi}{12\b^2} e^{ 2 a \mc{V}} +  \q_{\mathcal V \mathcal V} = \f{\pi}{12\b^2} e^{ 2 a \mc{V}}~,\nonumber
 \end{eqnarray}
while $\braket{T_{\mc{UV}}}_{\b}$ vanishes trivially (see Appendix \ref{App5} for detail derivation of these expressions including the normal ordered expressions $\braket{:T_{\mc U\mc U}:}_\b$ and $\braket{:T_{\mc V\mc V}:}_\b$). 
These are the Renormalized stress-tensor components for the scalar field immersed in a thermal bath, when seen from an accelerating frame of reference. We see that in the standard limits, either $a \to 0$ or $\b\to \infty$ leads us to the expected standard results. By this we mean that upon taking the $a \to 0$ limit, we recover the expected result for the stress tensor of a scalar field in thermal bath; and the limit for $\b \to \infty$, gives us vanishing components which is expected at zero temperature in flat spacetime. 
The first component of eq. (\ref{Tuu-TR}) represents the outgoing flux whereas the later one is related to ingoing flux. 

Let us now concentrate on the first part of the above equations which are arised from the normal order expansion. We see that these not only contain the standard factor of $a^2/48 \pi$, but also contain a spacetime dependent term. The last terms in these (i.e. second terms of the above equations) are arisen due to the already existing particle in the thermal bath. This can be understood in the following way. In Minkowski frame in presence of thermal bath we have $\braket{:T_{UU}:}_\beta = \pi/(12\beta^2)$. With respect to the accelerated frame this will be transformed to the second term of the above equation which can be checked by using tensorial transformation of  $\braket{:T_{UU}:}_\beta$ under Rindler transformation of coordinates. This tells that the first terms are purely due to the Unruh effect. 
The appearance of this spacetime dependence is due to the lack of time translational invariance of the Green's function eq. (\ref{Wight-TR}) which again signifies the non-equilibrium situation of the system. Moreover, the first expression of the above does not have any resemblance with the particle number (\ref{NTHR}) which one should expect. It implies that the application of the equilibrium statistical method in preceding subsection is not justifiable.

Before finishing this section, let us mention about the detector response for this case. It has been discussed in the previous subsection that the Wightman function is not time translational invariant in the proper Rindler frame. Therefore, one can not use (\ref{ResponseDefn}) for analytical discussion of detector response since it crucially depends on the time translational invariance of the Wightman function. This is also pointed out in \cite{Costa:1994yx}.  Of course, one can study the finite time detector response function (\ref{detc}) in this case. But we leave out this discussion in this paper.

\section{Comparison with Rindler-Rindler case} \label{Rind-rind}
In the previous section, we considered a Rindler observer moving through the thermal bath. In literature, it has already been established that the spectrum as perceived by a Rindler observer, is the same as that of a thermal bath. To rigorously establish this, one must compute proper observables and show that they give physically consistent results. With this spirit we also want to investigate how far such an indistingusibility between real thermal bath and quantum fluctuations in non-inertial frame exists. For that let us now compare the thermal-Rindler situation with Rindler-Rindler case. In this section, the Rindler-Rindler case will be discussed.

The Rindler-Rindler trajectory deals with an acceleration ``on top of another acceleration''. By this, we mean give another set of transformations in (\ref{Rtrans}) of the form:
\begin{eqn}
\mathcal{T} &= \f{e^{a_2 x}}{a_2} \sinh(a_2 t) \\
\mathcal{X} &= \f{e^{a_2 x}}{a_2} \cosh(a_2 t) 
\label{RRtrans}
\end{eqn}
Note that we have used a different value of acceleration for both the two separate transformations, $a_1$ and $a_2$. This leads to the following form of the metric: 
\be
\,d{s^2} =  e^{2\f{a_1}{a_2}e^{a_2 x} \cosh(a_2 t)} e^{2 a_2 x} (-\,d{t^2} + \,d{x^2})~,
\label{PRD2}
\ee
where the acceleration $a_1$, takes us from Minkowski to Rindler, and $a_2$ takes us from Rindler to Rindler-Rindler. Evaluating the two point correlation of the Rindler-Rindler fields squeezed between the Minkowski vacuum, we get, 
\be
\braket{\phi(u_2, v_2) \phi(u_1, v_1)}_M = \braket{\phi(u_2) \phi(u_1)}_M + \braket{\phi(v_2) \phi(v_1)}_M
\ee
Here, ($u,v$) are the Rindler-Rindler null coordinates, defined as 
\begin{equation}
u=t-x; \,\,\ v=t+x~,
\end{equation}
and $\braket{\cdots}_M$ denotes the expectation value taken between the Minkowski vacuum. The values of the quantities on the right hand side are given by, 
\begin{eqn}
 &\braket{\phi(u_2) \phi(u_1)}_M \\ 
 &\qquad =-\f{1}{4\pi} \log\Biggl[  \f{1}{a_1} \l\{ e^{\f{a_1}{a_2} e^{-a_2 u_1}} - e^{\f{a_1}{a_2} e^{-a_2 u_2}}\r\} \Biggr];\\
  &\braket{\phi(v_2) \phi(v_1)}_M \\ 
 &\qquad =-\f{1}{4\pi} \log\Biggl[  \f{1}{a_1} \l\{ e^{\f{a_1}{a_2} e^{a_2 v_2}} - e^{\f{a_1}{a_2} e^{a_2 v_1}}\r\} \Biggr]~.
\label{Wight-RR}
 \end{eqn}
 The above expressions are obtained by using two consecutive Rindler transformations in (\ref{G01}). 
Thus, from eq.(\ref{Wight-TR}) and eq.(\ref{Wight-RR}), one can easily observe that, at the level of the Green's function, the Rindler-Rindler case is very different from that of the Thermal-Rindler. To perform an explicit check of the equivalence of the two systems, one must compare some of the observables measured in the theory. The simplest of these include the Number operator, the Flux and the Response function. However, we easily see that the Green's function obtained in (\ref{Wight-RR}) is also not time translationally invariant. Thus, we again face similar issues like in the case of the Thermal-Rindler Green's function.  Thus, for comparison we compute the Flux of the outgoing modes of the scalar field in this spacetime and then compare it against eq.(\ref{Tuu-TR}).

From eq.(\ref{ST-general:2D}), we can obtain the value of the renormalized Stress-Tensor of the outgoing and in-going modes as, 
\begin{eqnarray}
\braket{T_{u u}(u)}_M &&=  \braket{:T_{u u}(u):}_M + \q_{uu}\nonumber\\
&&=\f{a_2^2}{48\pi} + \f{a_1^2}{48 \pi} e^{- 2 a_2 u} + \q_{uu}=0 ~;
\nonumber
\\
\braket{T_{vv}(v)}_M &&=  \braket{:T_{vv}(v):}_M + \q_{vv}  \nonumber\\
&&=\f{a_2^2}{48\pi} + \f{a_1^2}{48 \pi} e^{2 a_2 v} + \q_{vv}=0~.
\label{TuuTvv}
\end{eqnarray}
where $\q_{uu}$ and $\q_{vv}$ are given by
\begin{eqn}
    \q_{uu} = - \frac{a_1^2 e^{- 2 a_2 u} + a_2^2}{48\pi}, \quad 
    \q_{vv} = - \frac{a_1^2 e^{ 2 a_2 v} + a_2^2}{48\pi}~.
\end{eqn}
These are calculated from the expression ($\ref{PRD1}$) for the metric ($\ref{PRD2}$), expressed in null coordinates. 
The one that we have  chosen preserves the covariance of the stress tensor, as described in the previous section, and cancels with the other term in the equation above, resulting in the final renormalized value to be vanishing. A detailed discussion is given in Appendix \ref{App5}.
It is clear that if one recognises $a_1 = \f{2\pi}{\b}$ and $a_2 = a$, we exactly get the same result for the normal ordered stress tensor components as given in eq.(\ref{Tuu-TR}). However the complete renormalized stress tensor doesn't match as expected. This is an important consistency check, because it is not immediately obvious from the structure of the two Green's functions (eq.(\ref{Wight-TR}) and eq.(\ref{Wight-RR})), that they would lead to the same normal ordered stress tensors. Here, like earlier, appearance of two terms  is due to the same reason. The first Rindler frame sees the Minkowski vacuum as thermal which with respect to the second Rindler frame transforms to the last terms of the above expressions. While the first terms appear as if the second frame has an acceleration $a_2$ with respect to the first one.

The breakage of time translational invariance in the system, on going to the proper frame of the Rindler-Rindler observer (i.e, $u_1 = v_1 = \t_1, \ u_2 = v_2 = \t_2$) can be seen more explicitly by computing the Bogoliubov coefficients for the Minkowski to Rindler-Rindler transformation.
Using the Bogoliubov coefficients, one can evaluate the expectation values, 
\begin{eqn}
\braket{c_{q} c_{r}}_M, \ \braket{c_{q}^{\dagg} c_{r}^{\dagg}}_M, \braket{c_{q}^{\dagg} c_{r}}_M, \braket{c_{q} c_{r}^{\dagg}}_M,
\label{RR-Bogo-obs}
\end{eqn}
squeezed between the Minkowski vacuum. Here $c$ and $c^{\dagg}$ depicts the annihilation and the creation operator corresponding to the Rindler-Rindler observer, respectively. The last two in eq.(\ref{RR-Bogo-obs}) denote the number operator, and can be evaluated, as done in \cite{Kolekar:2013hra}. In the standard case, when one deals with Minkowski to Rindler transformation, it is found that $\braket{c_p c_q}_M$ and $\braket{c_p^{\dagg} c_q^{\dagg}}_M$ disappear for positive frequencies. However, it is seen that in the Rindler-Rindler case, this is not the case (for explicit expressions of these, see Appendix \ref{App6}). This distinctly shows that there is a breakage of time translational invariance in the system.

We evaluate the components of the stress-tensor in both cases, directly by using coordinate transformations and not relying on the step-wise evaluation of the Bogoliubov coefficients. Interestingly the renormalised  components do not match and hence the conjecture of equivalence is not valid at this level of investigation. However, it has been observed that first parts of (\ref{ST-general:2D}) for both the situations are same with the identification $a_1=2\pi/\beta$ and $a_2=a$. These terms can be interpreted as the stress-tensor components for an accelerating plane conductor. The same interpretation has been adopted earlier in \cite{Dowker:1978aza, Dowker:1994fi} for an accelerated conducting plane on a Minkowski spacetime. Therefore it must be noted that from this point of view, this equivalence conjecture is well satisfied, although the renormalised ones are not same.


\section{Rotating observer in thermal bath} \label{rot-therm}
So far we have concentrated on thermal-Rindler and Rindler-Rindler situations to investigate the issue of indistinguishability. This topic again will be explored in another very popular model. We shall here compare the rotating frame in the thermal bath with the Rindler-rotating frame. For that let us start calculating several observable quantities in this section for the case of thermal-rotating frame. Here we found that the detector response and number of particles can be calculated. The renormalised components of energy-momentum tensor are in principle calculable, but it turns out that we are not getting any readable expressions due to their huge structure (not even the package in {\it Mathematica $10$} can simplify them). Moreover, since the Rindler metric in four dimensions is not conformally flat, we can not use the existing results to obtain them. Therefore we leave this item in this paper. But it can be inferred that the components of renormalised stress-tensor will be non-vanishing due to presence of real thermal bath and the contribution due to coordinate transformation must vanish by the ``vacuum polarisation'' part as these are tensorial objects. This is similar to the thermal-Rindler case. In the next section the Rindler-rotating situation will be discussed.

\subsection{Detector response}
Although in literature there exists discussions \cite{Costa:1994yx, Hodgkinson:2014iua} on thermal-rotating case, here we shall investigate it in a much more detailed way so that ultimately it servers our main purpose. In this process some of the old results will again be investigated, but in such a way that it will be helpful to shed some light to our present aim. In addition, some new aspects of this model will be explored whenever necessary. It will be observed that the response function can be calculated by conventional way (i.e. using Eq. (\ref{ResponseDefn})) as the time translational invariance exists in the Green's function when expressed in rotating frame. We investigate the current topic in Cartesian coordinates in rotating frame. Analysis in Cylindrical coordinates is also presented in Appendix \ref{Cy} for completeness.

{\subsubsection{Rotating frame in Cartesian coordinates}} 
The time-like Killing vectors $\xi^a$ which generate rotating circular motion with respect to Minkowski spacetime is given by \cite{Gutti:2010nv}, 
\be
\xi^a = (\g, - \g \O Y, \g \O X, 0)~,
\ee
where $\Omega$ is angular velocity of the detector and the term $\g = (1 - \s^2 \O^2)^{-1/2}$ is called the Lorentz factor. The integral curve, in terms of rotating frame proper time, obtained from the above Killing vector, turns out to be \cite{Gutti:2010nv}:
\be
\tilde{x}(\t) = \big[ \g \t, \ \s \cos(\g \O \t), \ \s \sin(\g \O \t), \ 0 \big]~. 
\label{rot1}
\ee
Here, $\s $ is the radius of the circular path. Using these, we have 
\begin{eqnarray}
 &\D T = \g (\t_2 - \t_1) \equiv \g \breve{u}~, 
 \nonumber
 \\
 &|\D {\bf{X}}| = 2 \s \sin\l[ \f{\g\O (\t_2 - \t_1)}{2} \r] \equiv 2 \s \sin\l[ \f{\g \O \breve u}{2} \r]~.
\end{eqnarray}
 Substitution of the above in (\ref{GCar}) yields the positive frequency Green's function in rotating frame: 
\begin{eqn}
 &G_{\b}^+(\breve u) \\
 &= \frac{\mbox{cosec}\left( \frac{\O \g \breve u}{2} \right)}{16\pi \s \b}
  \Biggl(   \coth\l[ \f{\pi}{\b} (\g (\breve u-i\epsilon) + 2 \s \sin(\g \O \breve u/2)) \r] \\
 &\qquad\qquad\qquad\quad- \coth\l[ \f{\pi}{\b} (\g (\breve u-i\epsilon) - 2 \s \sin(\g \O \breve u/2)) \r] \Biggr)~.
\label{Green11} 
\end{eqn}
Note that the above is time translational invariant as it depends only on $\breve u$.
Upon taking the limit $\O \to 0$, this reduces to the familiar result for the Green's function of a stationery thermal bath, 
\be
\lim_{\O \to 0}G_{\b}^+(\breve u) = -\f{1}{4\b^2} \f{1}{\sinh^2\l[ \f{\pi}{\b}\g (\breve u - i \e) \r]}~.
\ee
The regulator $\epsilon$ in the above expressions indicates a choice of the positive frequency Green's function. 

 Substituting the Green's function  (\ref{Green11}) in (\ref{ResponseDefn}), we get the response function for an uniformly rotating detector in a thermal bath. Such an integral is not doable analytically and hence, we resort to using a numerical estimate. We also give another expression for the response function in cylindrical coordinates in Appendix \ref{Cy} which is sometimes very useful in certain situations. But here our whole analysis will be done by using (\ref{Green11}).

\subsubsection{Numerical analysis}
It's time to calculate the transition probability rate $\mathcal{R}(E)$ for the rotating observer in thermal bath. We have got the detector response function for this case after substituting the thermal Green's function (\ref{Green11}) into Eq.(\ref{ResponseDefn}). However, it does not seem to be possible to evaluate it analytically. Therefore, we solve the involved integration numerically to understand the features. For that we use the {\it Mathemetica} $10$ package.

\begin{figure}[h!] 
\centering
\scalebox{0.40}{\includegraphics[height=11cm,width=\textwidth]{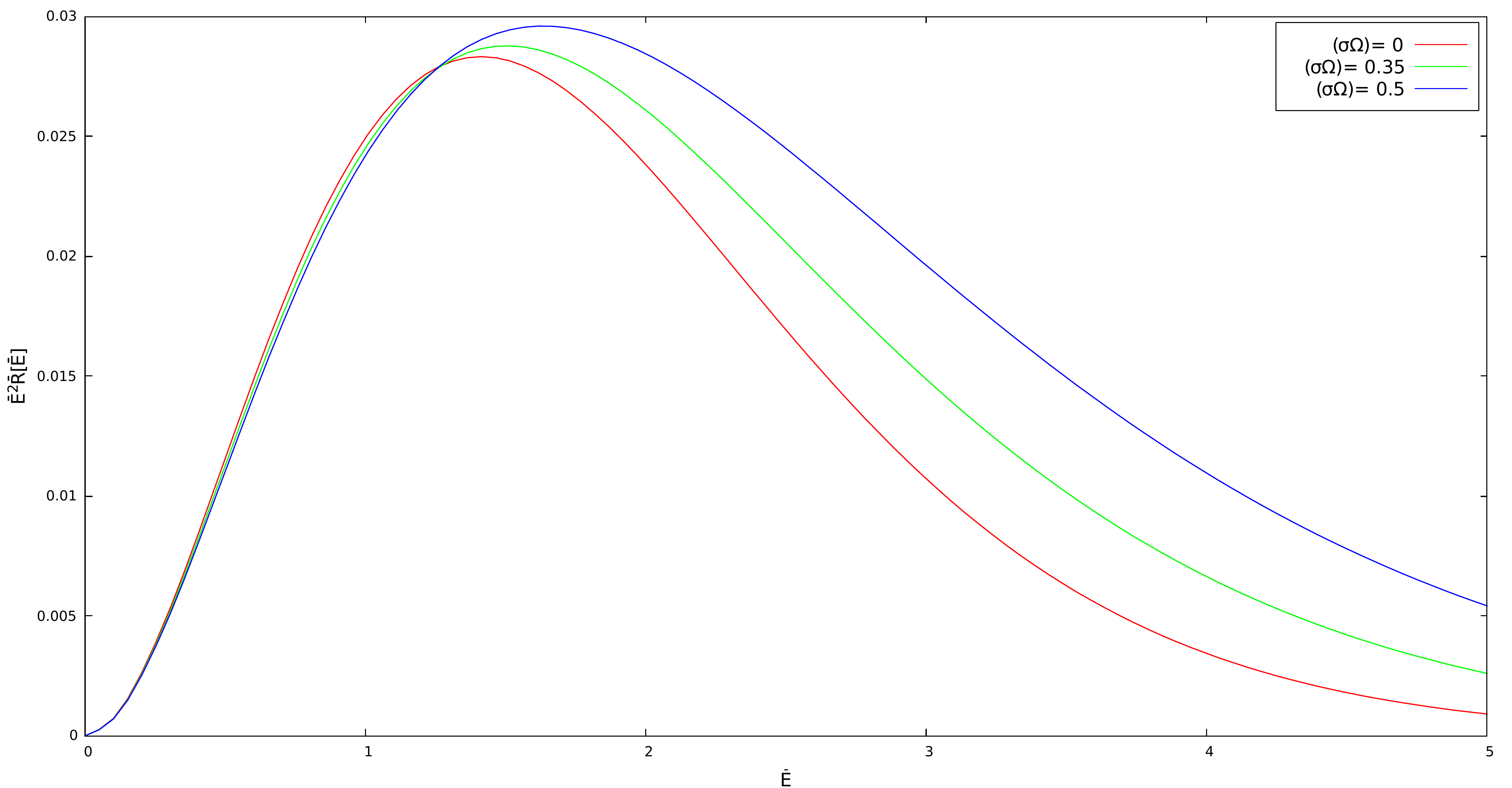}}
\caption{The response function of a uniformly rotating detector in cartesian coordinates has been plotted. To analyze the results we have chosen 
  two different values $({\s~\O})=0.35$, $0.5$ respectively
  and then compared with the response of static case ($({\s~\O})=0 $). We fix inverse temperature ${\b}=2$ and $\epsilon=0.01$. }
\label{Fig:1}
\end{figure}

Here we define the dimensionless energy which is $\bar{E}\equiv E/\g \O$.  Using this one finds that the dimensionless response function $\bar{\mathcal{R}}(\bar{E})\equiv \s \mathcal{R}(\bar{E})$ depends only on the dimensionless quantity $\s\O$. In Figure (\ref{Fig:1}) we plot the response function $\bar{E}^{2}\mathcal{\bar{R}}(\bar{E})$ with the variation of $\bar{E}$ for different values of $\s\O$. We shall apply these notations in all the plots through out our paper.  
\noindent
One can see from Figure (\ref{Fig:1}) that the distribution is similar to Planck.  Note that the peak increases as the angular velocity of the detector increases. Moreover, the rotating detector has always greater response than the static one.  This can be reconfirmed by the following analysis.

In the next Figure (\ref{Fig:3}) we plot the ratio of the response function for an inertial detector and an uniformly rotating detector.
\begin{figure}[h!] 
\centering
\scalebox{0.40}{\includegraphics[height=10cm,width=\textwidth]{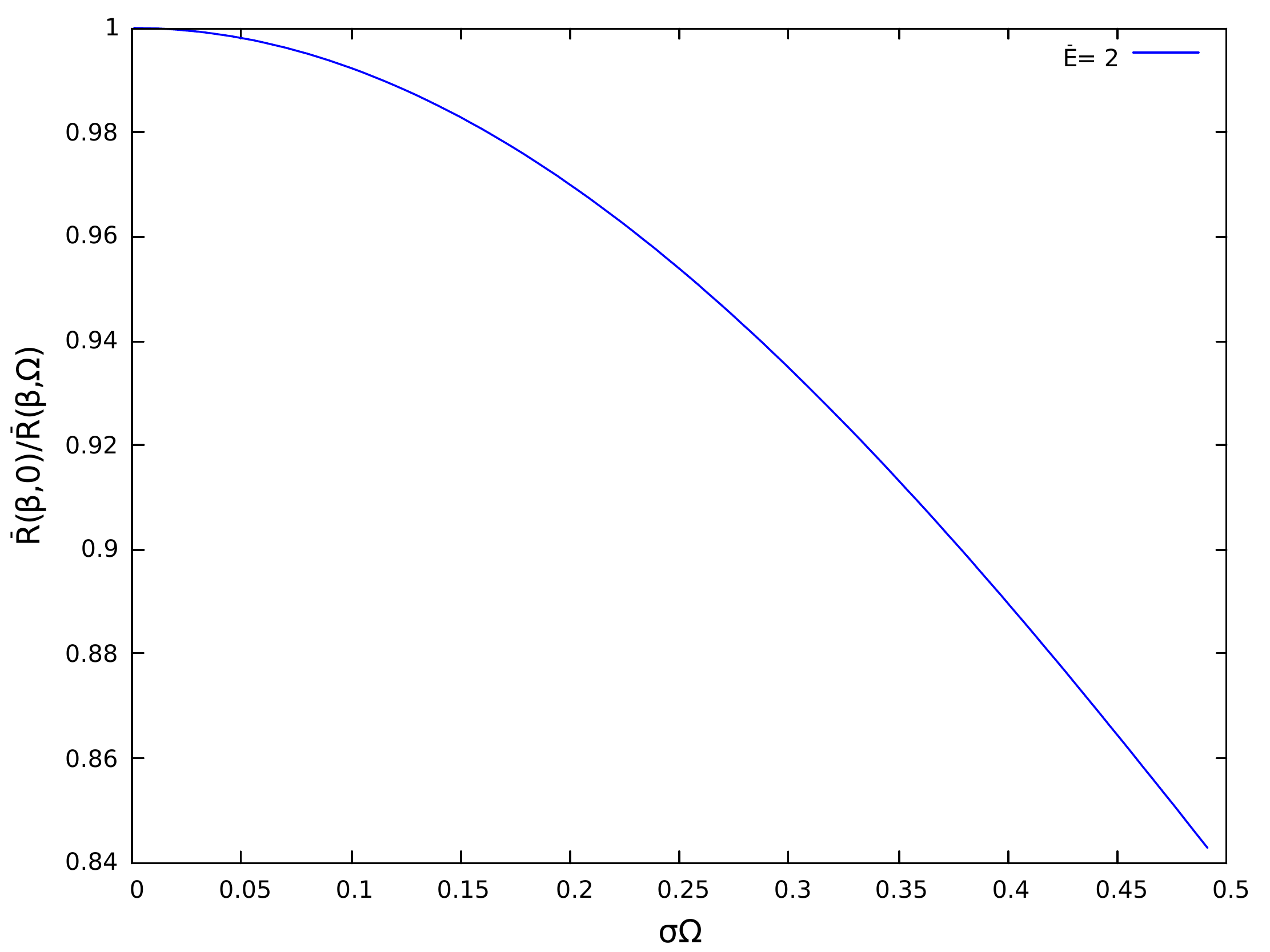}}
\caption{The ratio of response function between inertial($\O=0$) and uniformly rotating detector in real thermal bath of low temperature.
  The response of rotating detector is almost equal with static case in non-relativistic regime i.e $({\s~\O})\rightarrow 0$, but dominating in ultra-
  relativistic regime i.e $({\s~\O})\rightarrow 1$ . Here we take $\beta=2$ and $\epsilon=0.01$}
\label{Fig:3}
\end{figure}
\noindent
Here we can see that the response of thermal rotating detector is almost equal with thermal static case when ${\s \O}$ has lower value, but it starts dominating with the increase of $\s\O$. This concludes that they are both equal in the non-relativistic regime i.e $\s\O\rightarrow 0$ but in the ultra-relativistic regime thermal rotating response function dominates over the static case.

\begin{figure}[h!] 
\centering
\scalebox{0.45}{\includegraphics[height=10cm,width=\textwidth]{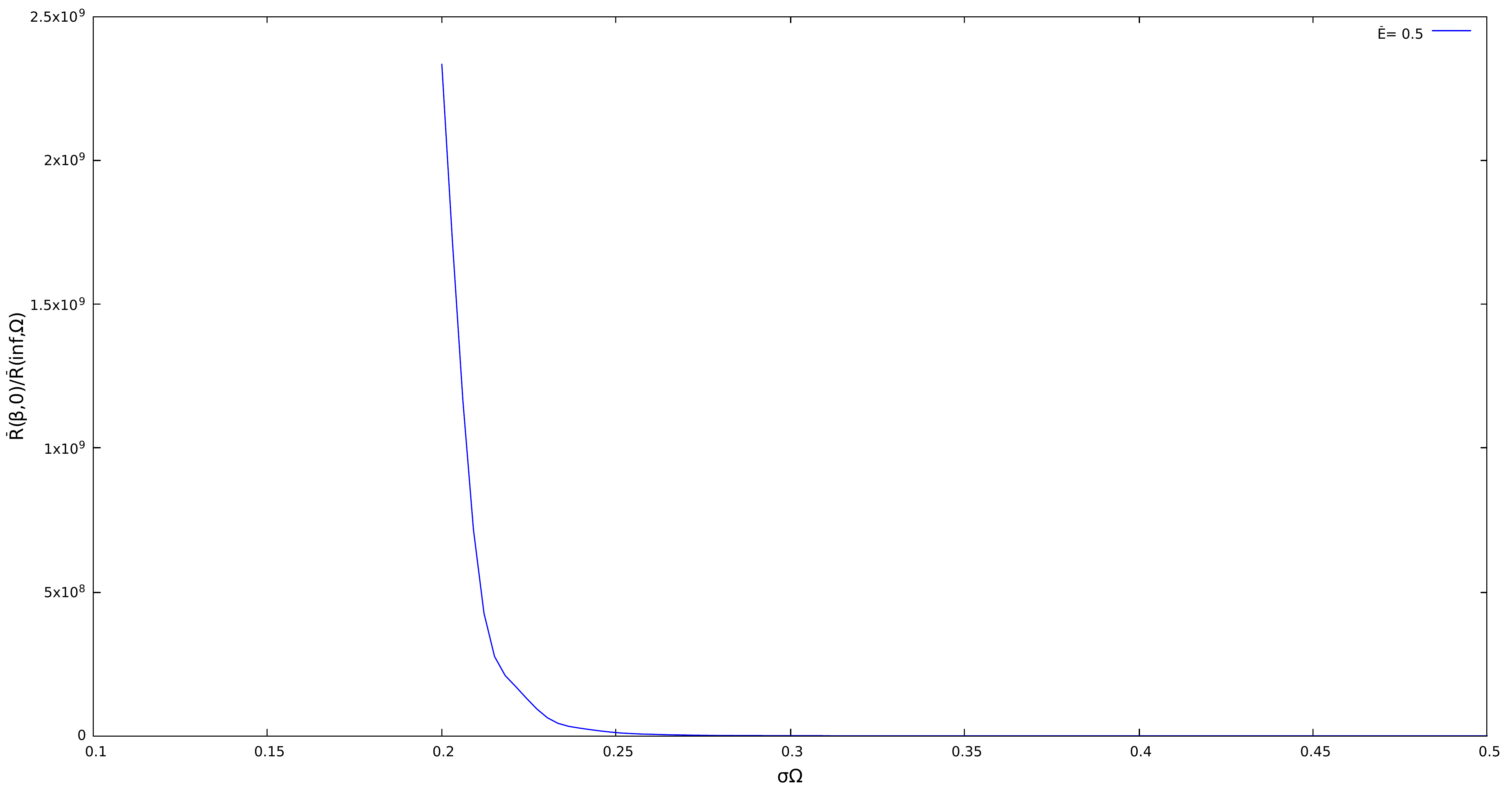}}
  \caption{The ratio of response function between static detector ($\O=0$) with finite temperature ($\beta=2$) and uniformly pure rotating detector ($\b\rightarrow \infty$).
  The response of rotating detector is always greater than thermal static one and it highly dominates in ultra-relativistic regime. The value of the regulator $\epsilon=0.01$.}
\label{Fig:4}
\end{figure}

Now we plot the ratio of the response functions for a static detector in a finite temperature background and a uniformly rotating detector with zero temperature at its background in Fig. \ref{Fig:4}.
\noindent
Here we can see the detector response for a pure rotating detector always dominates the thermal static one in ultra-relativistic regime. But the later one dominates for small value of $\s\O$. So, the ratio sharply falls with the increase of angular velocity. Although this discussion is not needed for our purpose, but for the sake of completeness of this section we present this one. 

In the above we numerically studied different features of the detector response function for the case of a rotating detector in a real thermal bath. Also we studied its properties at other limiting cases. The sole purpose of this analysis is not only understanding the response function, rather we compare this features with those for the Rindler-rotating model to understand how similar is the real thermal bath and the thermal bath seen from accelerating frame with respect to the rotating one. This will be done in the next section. Therefore, although it seems that the above graphical analysis contain a ``limited'' information; rather it actually gives us a very important massage for the paradigm of mimicking thermal bath by a non-inertial observer. It will help us to understand how far one can use the accelerated frame as a ``proxy'' for a real thermal bath. This topic is very important to understand the ``Unruh effect'' and people are investigating this issue for different situations (for example, see \cite{Kolekar:2013aka, Kolekar:2013xua, Kolekar:2013hra}). Our present analysis is completely in this direction.

\subsection{Particle production}
Now we want to calculate the particle number as measured by the rotating frame in the thermal bath.
This can be easily evaluated by relating the operators in the rotating frame to that of the Minkowski frame \cite{Paddy1}. Denoting the annihilation and the creation modes in the Rotating frame of reference by $\tilde{a}, \tilde{a}^{\dagg}$ respectively, and using the relations between these with those for the static observer, we find that the value of the number operator in the thermal bath becomes, 
\be
\braket{\tilde{a}^{\dagg}_i \tilde{a}_i}_{\b} =  \braket{{a}^{\dagg}_i {a}_i}_{\b}
\ee
Here, the index $i$ heuristically denotes the summation over all the momenta modes, as we need in the case of the number operator. The quantity $\braket{{a}^{\dagg}_i {a}_i}_{\b}$ denotes the thermal expectation value for the number operator of a free scalar field with respect to the static observer. This takes the form of the well known Bose-Einstein factor $\sim \bose_{\b}(i)$.

Let us now make some comment on this result. The obtained result clearly shows that there are no new particles produced in the case of a rotating detector, even when it is placed in the thermal bath. Also it may be noted that the response function of this system, obtained in the previous section, behaves differently from this number operator. We just observed that the response function increases with the increase of rotational velocity of the detector, which clearly indicates that the rotation in the observer affects its detection. Whereas, the number of particles detected, calculated by number operator, does not depend on the rotational parameter.

In general, there is no reason to expect that the response function, obtained here, will be equal to the Number operator's expectation value. It is a coincidence, that in even dimensional cases and for an uniformly accelerating detector, they equate to the same value and give an indication for the thermal spectrum. If one considers a detector moving in a different trajectory, for example, an uniformly rotating trajectory, there is a conflict between the number of particles detected (vacuum expectation value of the Number operator) and the response function \cite{Paddy1}. For the case of a uniformly rotating detector, we get a non-zero value for the the response function, but the number of particles evaluates to zero. This clearly demonstrates that the uniformly rotating detector does not detect any real particle, but does have a finite response. To better understand this scenario, and to see that the rotating detector truly never registers any new particle, her we probe a thermal scalar bath with such a uniformly rotating detector. 
 
\section{Comparison with Rindler-Rotating case} \label{rind-rot}
In the previous section we calculated the response function for the rotating observer in thermal background. Now in this section we shall demonstrate the calculation for a rotating observer in a Rindler frame in cartesian coordinates. Our aim is to compare the detector's response between the thermal-rotating and the Rindler-rotating case and finally draw a conclusion.  As we shall notice that the system is completely in non-equilibrium we shall not calculate the number of particles here. Also we find difficulty to calculate the components of renormalised stress-tensor for their huge structure (even package in {\it Mathematica} $10$ fails to do that) and so we leave this for the moment. But it can be agued that all the renormalised components must vanish as the corresponding Minkowski values are zero.

\subsection{Detector response}
In order to calculate the detector response for the Rindler-rotating observer we need to calculate the Wightman function in Monkowski spacetime first, which is given by (\ref{G0}).

To get the expression of the Wightman function in Rindler-rotating frame first we have to get the form of space-time interval $(\Delta T^{2} - \vert{\Delta {\bf{X}}}\vert^{2})$ in the Rindler-rotating frame. In the prvious section we have got the form of integral curve in terms of rotating frame proper time from a time-like Killing vector $\xi^{a}$ which generate rotation with respect to Minkowski spacetime (see Eq.(\ref{rot1})). Now applying this rotational transformation on the space-time interval in Rindler frame we have
\begin{eqn}
&\Delta T^{2} - \vert{\Delta {\bf X}}\vert^{2}\\
&=-\frac{1}{a^{2}}\left(e^{2a\sigma\cos{\gamma\Omega\tau_{1}}}+e^{2a\sigma\cos{\gamma\Omega(\breve u+\tau_{1})}}\right) 
\\
&\quad +\left(\frac{2}{a^{2}}\ e^{2a\sigma\cos{\frac{\gamma\Omega(\breve u+2\tau_{1})}{2}}\cos{\frac{\gamma\Omega\breve u}{2}}}\right)\cosh{a\gamma(\breve u - i\epsilon)}
\\
&\qquad - 4\sigma^{2}\cos^{2}{\frac{\gamma\Omega(\breve u+2\tau_{1})}{2}}\sin^{2}{\frac{\gamma\Omega \breve u}{2}}~.
 \label{rindrot}
\end{eqn}
This procedure, however creates a problem as to the definition of the Wightman function. As one can check that the Wightman function for the Rindler-rotation case which we have got is not invariant under the time translation. Therefore we need a prescription which is to compute the value of the detector response for a finite time interval. In that case we need to replace our Wightman function by what we shall call the {\it regularized Wightman function} which is defined by
\begin{eqnarray}
\mathcal{W}_{R}(\tau_{1}, \breve u)=-\frac{1}{4\pi^{2}}\left[\frac{1}{\Delta T^{2} - \vert{\Delta {\bf X}}\vert^{2}}-\frac{1}{(\breve{u} - i\epsilon)^{2}}\right]~,
\label{reg}
\end{eqnarray}
with the response function is given by (\ref{detc}).
As it is explained in \cite{Barbado:2012fy} that this quantity (\ref{reg}) is  well defined as a function where the pole at $\breve u=0$ can be avoided by subtracting the extra factor (A similar regularised Wightman function for ($1+1$) dimensional case has been advocated in \cite{Juarez-Aubry:2014jba} for a derivative type coupling).

\begin{figure}[!h]
\centering
\scalebox{0.45}{\includegraphics[height=11cm,width=\textwidth]{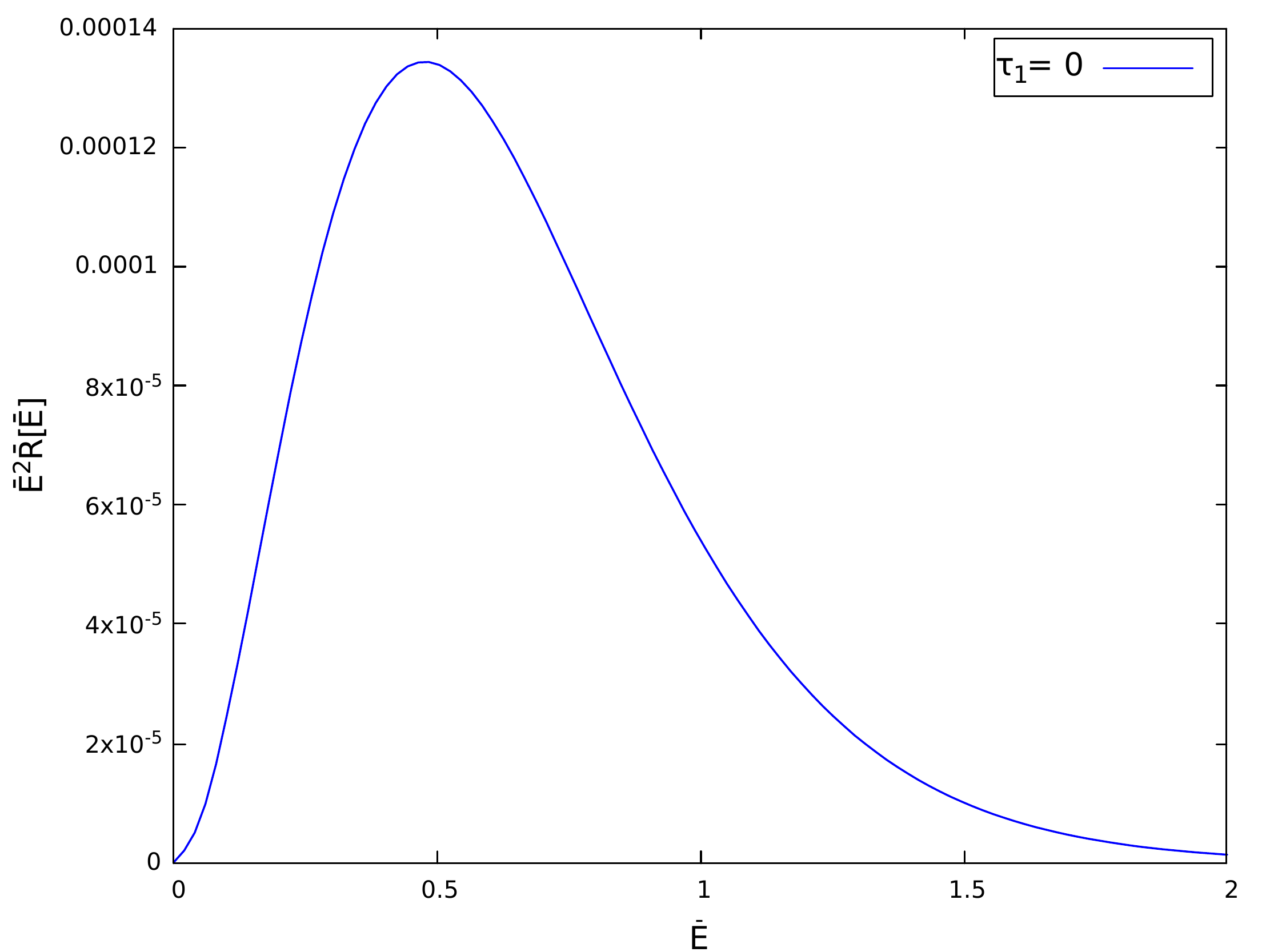}}
\caption{The response function of a uniformly rotating detector in Rindler coordinates versus Energy has been plotted. Taking the initial proper time $\tau_{1}=0$, $\sigma\Omega=0.15$, $\sigma a=1$ and $\epsilon=0.5$.}
\label{Fig:5}
\end{figure}

We solve this integration (\ref{detc}) numerically for a finite time interval taking the initial proper time $(\tau_{1})$ to be zero and then plot it. This is presented in figure \ref{Fig:5}.
\noindent
Interestingly, we can see from Figure \ref{Fig:5} that the response function of a Rindler rotating detector have the Planck distribution for a finite time which is analogous to the response of a rotating detector in a thermal bath (see Fig. \ref{Fig:1}). This implies that both thermal-rotating and Rindler-rotating behaves in a similar way.

\begin{figure}[!h]
\centering
\scalebox{0.45}{\includegraphics[height=11cm,width=\textwidth]{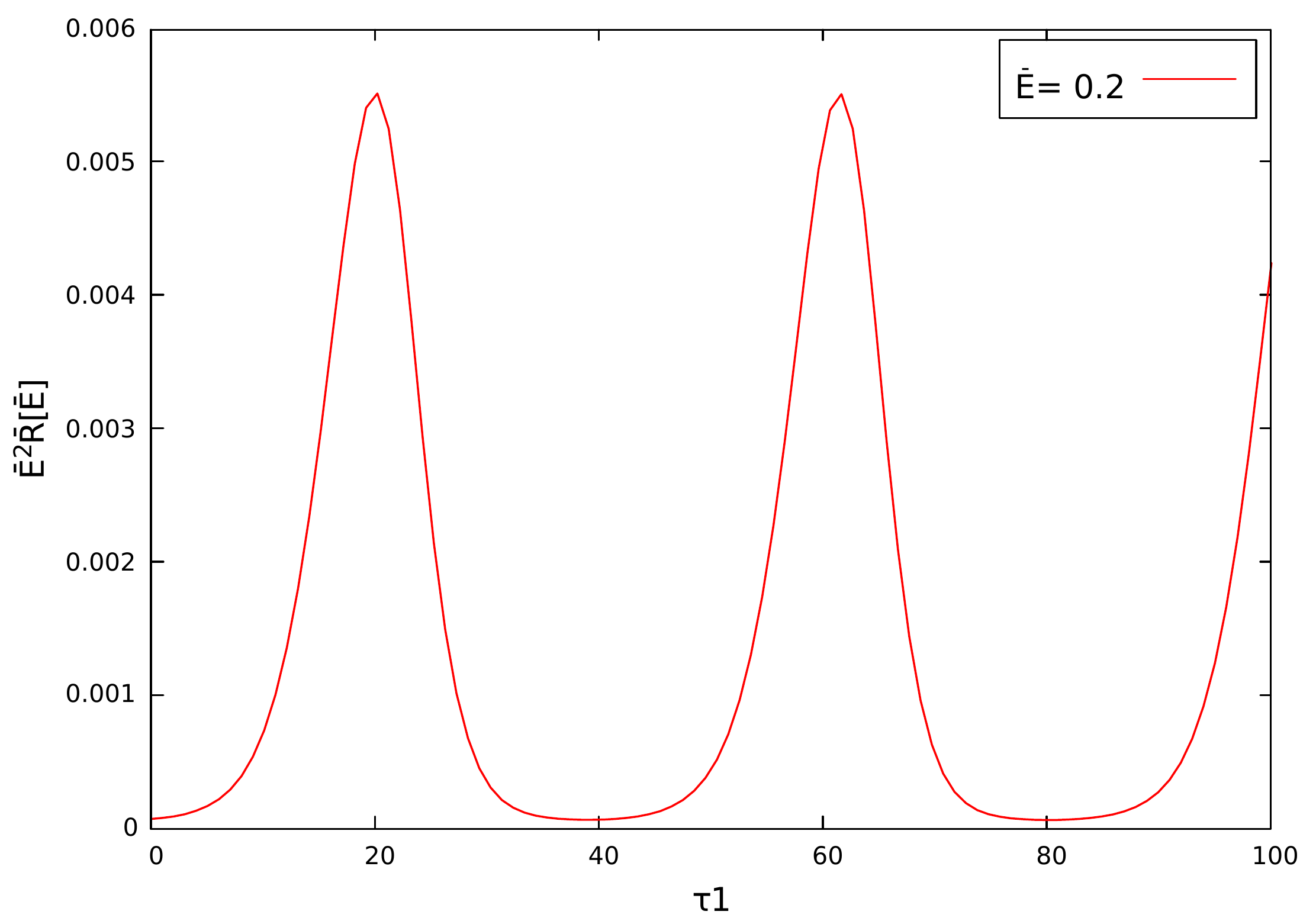}}
\caption{The response function of a uniformly rotating detector in Rindler coordinates with respect to the initial time has been plotted by taking $\sigma\Omega=0.15$, $\sigma a=1$ and $\epsilon=0.5$.}
\label{Fig:6}
\end{figure}

However, there exists a crucial difference between the response functions in these two situations. This will be elaborated in following discussion. For that below we plot the response function with the variation of the initial proper time in Fig. \ref{Fig:6}. 
\noindent
We have found that the response function is reaching its peak value with a certain periodicity which is a unique observation not found in the case of the thermal rotation case. It can be shown that the frequency of getting the peak value by the response function is equal to the frequency of the rotation of the detector which shows our calculations are correct. One can check that the same is also evident from the analytical expression. 

Let us now explain the significance of our investigation and how it serves our main goal. We have studied different properties of the detector response function for two scenarios which are the thermal rotating model and the Rindler-rotating model. During the computation we have found that the that the Green's function for the thermal rotating observer is invariant under the time translation but not in the case for Rindler-rotating observer. As a result we performed the integration numerically for the detector response for the Rindler-rotating case within the finite time interval unlike the thermal rotating case where we performed the integration in the time limit from $-\infty$ to $\infty$. Next, we have performed different features of the detector response function numerically and analysed them with the graphical representations for both the cases. The information contained in each graph itself may seem to be very ``limited'' at a first glance by the reader but the result is interesting when one compare these two pictorial views (Fig. \ref{Fig:1} and Fig. \ref{Fig:6}) which shows a clear distinction between them.  But this distinction is not quite vivid to visualise by comparing the numerical values for both the cases. Therefore, the sharp contrast between these graphical analysis raises many questions about the indistinguishability between the real thermal bath and the thermal bath seen from the accelerating frame with respect to the rotating one. So this result is an important one as it resonates the fact that the indistinguishability
between the quantum fluctuation seen from non-inertial
observer and the real thermal bath itself may be an observer
dependent statement. It will help us to shed light on the fact that how far the accelerated frame can used as a ``proxy'' for a thermal field and the vice versa. Therefore we feel that the present graphical analysis plays an important role to understand the issue, which is under investigation in this paper.
\section{Summary and Conclusions} \label{conc}
Let us now summarise what we have done in this paper and discuss their implications. Our aim was to investigate the robustness of the indistinguishability between the quantum fluctuation seen by a non-inertial observer and the real thermal bath. For this purpose, we looked at the thermal bath and the Rindler frame in Minkowski spacetime from a non-inertial frame. Here two non-inertial observers have been selected: one is another Rindler observer and other one is a uniformly rotating observer. To obtain a comparative study, we computed different observables in these models and compared them. In the following we summarise our observations.

First concentrate on the thermal-Rindler and Rindler-Rindler model. This case has been studied earlier. But, as stated earlier, it is needed to be looked at again. Here we observed that at the level of the Green's function, the Rindler-Rindler case is very different from that of the Thermal-Rindler case and interestingly both of their Green's function are not invariant under time translation. So, the problem arises for evaluating the number of particles as it is dependent on the Green's function. In that case the better way is to calculate the expectation value of the stress tensor of the system which basically tells us the number of quanta emitted per unit area. Using the Green's function we obtained the value of components of the renormalised stress tensor and found that they differ. Although one cannot use this as a measure for distinguishing them experimentally, it does tell us that there is some difference between as scalar field living in Rindler space and it living in a thermal bath. However, at the level of normal ordered stress tensors, the two results exactly match.

Next we investigated the same in a different set up. Here thermal bath and the Rindler frame in Minkowski spacetime have been studied from a uniformly rotating frame.  Here again at the Green's function level we found these two cases are different. Although the Green's function for the thermal rotating observer is invariant under time translation but not in the case for the Rindler rotating observer. As a result we found that the detector response function for the Rindler rotating observer is dependent on the initial proper time, whereas for the thermal rotating observer it is independent of that. We plotted the response function for both the cases and found that the thermal-rotation and Rindler-rotation both give the standard Planck distribution. But there is an additional feature presents in the later case. The Green's function is not time translational invariant and there exists a certain periodicity in the response function which is absent in for thermal bath case as the system is in equilibrium.
       
To conclude, we mention that the thermal rotating case is time translational invariant whereas Rindler-rotating is not. Consequently detector response in first case do not show periodicity with time while the later one does show this. This is clearly a difference between these two. 
Whereas so far we see the Thermal-Rindler and Rindler-Rindler are equivalent in all aspects except the values of the components of the renormalised stress-tensor. Therefore it may be the case {\it the equivalence between real thermal bath and the Rindler frame is not totally guaranteed}.  Of course, this is not a conclusive statement, rather a suggestive one. In this regard, it must be mentioned that even the equivalence between the accelerated frame and the real thermal bath at the level of their own proper frames holds only in two and four dimensions (see the discussion below Eq. $(4.2.15)$ in section $4.2$ of \cite{R4}). Here we observed that even in these dimensions they are not quite similar with respect to a new set of non-inertial observers. We hope the present analysis shed some light in this particular issue.


\vskip 3mm
\noindent
{\bf{Acknowledgment:}}\\
\noindent
The authors like to thank Sanved Kolekar for clarifying his related papers and making several useful comments at the initial stage of the work. We also thank the anonymous Referee for bringing important references to our notice which helped us to improve our manuscript. 
\begin{appendix}
\section{\label{App4}Derivation of Eq. (\ref{Graw})}
Consider the expansion of a scalar field $\phi(T,\textbf{X})$ in terms of creation and annihilation operators,
 \begin{equation}
  {\P}(T,{\bf {X}})=\sum_n\frac{f_n(\bf {X})}{\sqrt{2 {\o}_n}}\Big (a_n e^{-i{\o}_nT}+a_n^{\dag}e^{i{\o}_nT}\Big)~.
  \label{scal}
 \end{equation}
The expectation value of two point function in thermal state is give by,
\begin{equation}
 G_\beta(X_2;X_1)={\la}{\P}(X_2){\P}(X_1){\ra}=\frac{1}{Z} {\textrm{Tr}}\Big[e^{-{\b}H}{{\P}(X_2){\P}(X_1)\Big]}~.
 \label{tpf}
\end{equation}
Now since the scalar field is a collection of infinite number of Harmonic oscillators, to compute the above we choose $H=a^{\dagg}a\omega_n$ with $a$ ($a^{\dagg}$) is the annihilation (creation) operator. To take the trace, we choose the energy eigenstates of the Harmonic oscillator. Here $Z$ is the partition function which is given by
\begin{eqn}
 Z&={\textrm{Tr}}(e^{-\b H})
=\sum_{n} \bra{n}e^{-\b(a^{\dagg}a)\o_{n}}\ket{n}\\
 &=\sum_n e^{-n\b{\o}_n}=(1-e^{-\b{\o}_n})^{-1}~.
 \label{part}
\end{eqn}
Using this and the mode decomposition (\ref{scal}) in (\ref{tpf}) one obtains Eq.\eqref{Graw}.
\section{\label{App3}Finite temperature to zero temperature propagator in $2$-dimensions}
 The thermal Green's function in $2$-dimensions has the form as stated in eq. (\ref{G2D}). It can be seen that this gives back the conventional result for the Wightman function in $2$-dimensions for a massless scalar field at zero temperature. This result however, is not directly obvious, because at face it looks as if it diverges in the limit $\b \to \infty$. However, if the limit is taken carefully, then one can see that, upto the divergences which are encountered in a $2D$ massless scalar propagator, we get back the known result (\ref{G01}).
 
 The steps are as follows. First expand the exponential factor within the logarithmic function in (\ref{G2D}) and keep upto first order in $1/\beta$. This leads to
\begin{eqnarray}
G_\beta(X_2,X_1) =&&  -\frac{1}{4\pi}\Big[\ln \Big\lbrace\frac{2\pi}{\beta}(\Delta T - \Delta X)\Big\rbrace
\nonumber
\\
&&+\ln\Big\lbrace\frac{2\pi}{\beta}(\Delta T+\Delta X)\Big\rbrace\Big]~.
\end{eqnarray}
Now the finite term of the above is
\begin{equation}
 G_\beta(X_2,X_1) \sim  -\frac{1}{4\pi}\Big[\ln(\Delta T - \Delta X)(\Delta T+\Delta X)\Big]~,
\end{equation}
which in null-null coordinates transforms to Eq. (\ref{G01}).
\section{\label{App1}Derivation of Eq. (\ref{GCar})}
Eq. (\ref{1}) can be evaluated in the following way. Using the spherical polar coordinate representation of momentum coordinates ($k,\theta,\Phi$) and using $\omega=k$ we find:
\begin{eqn}
G_\beta(X_2;X_1) &= \int \f{k^2\sin\theta\ dk\ d\theta\ d\Phi}{(2\pi)^3} \f{e^{i k |\bf{X}| \cos{\q}}}{2 k}\\ 
&\qquad\qquad\times\f{1}{e^{\b k} - 1} \Bigl[ e^{i k T} + e^{\b k} e^{-i k T} \Bigr] \\
&= \int_0^{\infty} \f{\,d{k}}{(2\pi)^2} \f{k \Bigl[ e^{i k T} + e^{\b k} e^{-i k T} \Bigr]}{2 (e^{\b k} - 1)} \\
&\qquad\qquad\times\int_{-1}^{1}\,d({\cos{\q}}) e^{i k |\bf{X}| \cos{\q}} \\
&= \f{1}{(2 i |\bf{X}|)} \int_0^{\infty} \f{\,d{\o}}{(2\pi)^2} \f{1}{e^{\b \o} - 1} \\
&\qquad \times\Bigl( e^{i \o T} + e^{\b \o} e^{-i \o T} \Bigr) \Bigl( e^{i \o |\bf{X}|} - e^{-i \o |\bf{X}|} \Bigr) \\
&= \f{1}{2i|\bf{X}|} \sum_{n = 1}^{\infty} \int_0^{\infty} \f{\,d{\o}}{(2\pi)^2} e^{-n\b\o} \\
&\qquad \times\Bigl( e^{i \o T} + e^{\b \o} e^{-i \o T} \Bigr) \Bigl( e^{i \o |\bf{X}|} - e^{-i \o |\bf{X}|} \Bigr) \\
&= \f{1}{2i|\bf{X}|} \f{1}{(2\pi)^2} \mathcal{S}~,
\label{thermal prop1}
\end{eqn}
where
\begin{eqn}
 \mathcal{S} &= \sum_{n = 1}^{\infty} \underbrace{\f{1}{i(T - |{\bf{X}}|) + (n-1)\b} + \f{1}{i(T - |{\bf{X}}|) - n \b}}_{\mathcal{S}_1} 
 \\
 &\qquad +  \underbrace{\f{i}{(T + |{\bf{X}}|) + i n \b} + \f{i}{(T + |{\bf{X}}|) - i (n - 1)\b}}_{\mathcal{S}_2}~.
\end{eqn}
It can be shown that, 
\begin{eqn}
 \mathcal{S}_1 &= - \f{i \pi}{\b} \coth\l[ \f{\pi}{\b} (T - |\bf{X}|) \r]~; \\
 \mathcal{S}_2 &= \f{i \pi}{\b} \coth\l[ \f{\pi}{\b} (T + |{\bf{X}}|) \r]~.
\end{eqn}
Substitution of this in (\ref{thermal prop1}) yields (\ref{GCar}).

Although the form of Green's function (\ref{GCar}) is enough for our purpose, but for completeness we shall see that the same can be expressed in an another useful form. For this we rewrite $\mathcal{S}$ in (\ref{thermal prop1}) by using $\omega=k$ as 
\begin{eqn}
   \mathcal{S}   &= \intsinf \f{dk}{e^{\b k} - 1} \Bigl[ e^{ik(T + |\bf X|)} - e^{ik(T - |\bf X|)} \\
   &\qquad + e^{\b k} e^{-ik(T - |\bf X|)} - e^{\b k} e^{-ik(T + |\bf X|)} \Bigr] \\
   &= \underbrace{ \intsinf \f{dk}{e^{\b k} - 1} \Bigl[ e^{ik(T + |\bf X|)} - e^{ik(T - |\bf X|)} \Bigr]}_{I_{1}} \\
   & \underbrace{+ \int_{-\infty}^0 \f{dk}{e^{\b k} - 1} \Bigl[ e^{ik(T + |\bf X|)} - e^{ik(T - |\bf X|)} \Bigr]}_{I_{2}}
  \end{eqn}
 where in $I_2$, $k$ has been replaced by ($-k$).
Therefore one obtains
 \begin{eqn}
\mathcal{S}&= I_{1} + I_{2}\\
 &=\intinf \f{dk}{e^{\b k} - 1} \Bigl[ e^{ik(T + |\bf X|)} - e^{ik(T - |\bf X|)} \Bigr]~.
 \end{eqn}
 The above integration can be done by Complex integration technique.  This has simple pole at $k=(2i\pi n)/\beta$ with $n=0,\pm 1, \pm 2\dots$. Now since here we have $T-|\bf{X}|>0$ and $T+|\bf{X}|>0$ , the contour will in the upper half and hence one should have $k=(2i\pi n)/\beta$ with $n=1, 2\dots$. In this situation, the above leads to
 \begin{eqn}
  \mathcal{S}
  &=  -\f{2\pi i}{\b} \sum_{n = 1}^{\infty} \Bigl[ e^{- \f{2\pi n}{\b} (T - |\bf X|)} - e^{- \f{2\pi n}{\b} (T + |\bf X|)} \Bigr] \\
  &= -\f{2\pi i}{\b} \l[  \f{1}{e^{\f{2\pi}{\b}(T - |\bf X|) } - 1 } - \f{1}{e^{\f{2\pi}{\b}(T + |\bf X|) } - 1 } \r] \\
  &= -\f{4\pi i}{\b} \f{\sinh\l(  \f{2\pi}{\b} |\bf X| \r)}{\cosh\l( \f{2\pi}{\b} T \r) - \cosh\l( \f{2\pi}{\b} |\bf X| \r)}~.
\end{eqn}
Then the another form of Thermal Green's function in position space can be expressed as,
\begin{eqn}
G_\beta^+({X}_2,{X}_1)= -\f{1}{2\pi |\bf X| \b} \f{\sinh\l(  \f{2\pi}{\b} |\bf X| \r)}{\cosh\l( \f{2\pi}{\b} T \r) - \cosh\l( \f{2\pi}{\b} |\bf X| \r)}~.
\end{eqn}

\section{\label{App2}Derivation of Eq. (\ref{NTHR})}
One can express the creation and annihilation Rindler operators in terms of Minkowski operators using the Bogoliubov transformations as,
\begin{eqnarray}
b_{R}=\int_{0}^{\infty}ds \left(\alpha_{sR}a_{s} +\beta^{*}_{sR}a_{s}^{\dagg}\right)
 \\
 b_{Q}^{\dagg}= \intsinf \,d{p} \left(\a_{pQ}^* a_p^{\dagg} + \b_{pQ} a_p\right)~
\label{ab}
\end{eqnarray}
where the values of the Bogoliubov coefficients $\a$ and $\b$ for positive frequencies are given by,
\begin{eqnarray}
\a_{pQ} &= \f{\theta(p)}{2 \pi a} \sqrt{\f{Q}{p}} e^{\pi Q/ 2a} \l(\f{a}{p}\r)^{\f{iQ}{a}}  \Gamma\l(  \f{i Q}{a} \r)~;
\nonumber
\\
\b_{pQ} &=- \f{\theta(p)}{2 \pi a} \sqrt{\f{Q}{p}} e^{-\pi Q/ 2a} \l(\f{a}{p}\r)^{\f{-iQ}{a}}  \Gamma\l(  \f{-i Q}{a} \r)~.
\label{alpha}
\end{eqnarray}
For simplicity in notations we have distinguished the Rindler and the Minkowski frequencies by block letters and small letters. $a$ and $a^{\dagg}$ denote the Minkowski creation and annihilation operators respectively and $b$ and $b^{\dagg}$ denote the Rindler creation and annihilation operators. 
In order to evaluate equation (\ref{NB}) we let us first find the value of $ \braket{n|b_{Q}^{\dagg}b_{R}|n}$. This can be evaluated in the following way. Use of (\ref{ab}) leads to 
 \begin{eqn}
 \braket{n|b_Q^{\dagg} b_R|n} &= \intsinf \,d{p} \,d{s} \Bigl[  \braket{n|a_p^{\dagg} a_s|n} \a_{pQ}^* \a_{sR} \\
 &+ \braket{n|a_p a_s^{\dagg}|n} \b_{pQ} \b_{sR}^* \Bigr] \\ 
 &= \intsinf \,d{p} \Bigl[  (n) \a_{pQ}^* \a_{p R} + (n + 1) \b_{pQ} \b_{pR}^* \Bigr]~.
 \label{bb}
 \end{eqn}
Now  using (\ref{alpha}) one obtains
\begin{eqnarray}
 &\intsinf \,d{p} \ \a_{pQ}^* \a_{pR} = \f{e^{\pi Q/a}}{4\sinh\l( \f{\pi Q}{a} \r)} \ \d\l( \f{Q - R}{a} \r) ~;
 \\
 \label{alal}
&\intsinf \,d{p} \ \b_{pQ}^* \b_{pR} =  \f{e^{-\pi Q/a}}{4\sinh\l( \f{\pi Q}{a} \r)} \ \d\l( \f{Q - R}{a} \r)~.
 \label{bebe}
 \end{eqnarray}
 Then (\ref{bb}) reduces to the following form:
  \begin{eqn}
 \braket{n| b_Q^{\dagg} b_R |n} = \f{1}{2} \Biggl[  \f{n}{1 - e^{-\f{2\pi Q}{a}}} +  \f{n + 1}{e^{\f{2\pi Q}{a}} - 1} \Biggr] \d\l( \f{Q- R}{a} \r)~.
\label{nbb}
 \end{eqn}
 
 Next we re-express (\ref{NB}) as
 \begin{eqn}
  \braket{\mathcal{N}}_{\b}&= \f{1}{Z} \intsinf \f{\,d{P}}{2P}  \sum_{n,m= 0}^{\infty} \braket{n| b_{P}^{\dagg} b_{P}|m}  \braket{m| \exp[-\b a_{\o}^{\dagg} a_{\o}] |n} \\
  & = \f{1}{Z} \intsinf \f{\,d{P}}{2P}  \sum_{n= 0}^{\infty} \braket{n| b_{P}^{\dagg} b_{P}|n}  e^{-n\beta\omega}~. 
   \end{eqn}
which after substitution of (\ref{nbb}) leads to
  \begin{eqn}
 \braket{\mathcal{N}}_{\b}&= \f{1}{4Z} \intsinf \f{\,d{P}}{P}  \sum_{n= 0}^{\infty}   e^{-\b \o n} \Biggl[   (n )(1+\bose_a(P)) \\
  &+ (n + 1) \bose_a(P)\Biggr] \delta(0)\\
  &= \f{1}{4Z} \intsinf \f{\,d{P}}{P} \f{e^{\b\o}}{(e^{\b\o} - 1)^2} \Bigl[ \bose_a(P) (1 + e^{\b\o}) + 1  \Bigr]\delta(0) 
  \label{num}
  \end{eqn}
where we introduced notations like
 \be
 \bose_a(P) = \f{1}{e^{\f{2\pi}{a} P} - 1}, \quad \bose_{\b}(P) = \f{1}{e^{\b P}  - 1}
 \ee
 which have the following properties
 \begin{eqn}
  1 + \bose_{\b}(\o) = - \bose_{\b}(-\o), \quad e^{\b\o} \bose_{\b}(\o) = 1 + \bose_{\b}(\o)~.
 \end{eqn}
 Finally, using 
 \be
 Z = \sum_{n = 0}^{\infty} \braket{n| \exp[-\b a_{\o}^{\dagg} a_{\o}] |n} = \f{e^{\b \o}}{e^{\b \o} - 1} = 1 + \bose_{\b}(\o) 
 \label{par}
 \ee 
one obtains Eq.\eqref{NTHR}.
\section{\label{App5}Derivation of Eq. (\ref{Tuu-TR}) and (\ref{TuuTvv})}
The stress tensor is a composite operator, and upon the naive computation of its expectation value, it diverges. Thus, one must follow a well-defined regularization scheme in order to evaluate the expectation value of such quantities. One of the common techniques used, is the point-splitting method \cite{Davies:1976hi}. The point splitting method can be described in the following steps:
  \begin{enumerate}
   \item Evaluate the expectation value of the composite operator by imagining it to be a non-local object, i.e, a distribution across a few points.
   
   \item Carefully take the limit, where all the points (across which the correlator is distributed) tend to a single point. 
   
   \item Extract the divergence piece and the finite term, and identify the renormalized correlator. 
  \end{enumerate}
\noindent
  This series of steps are better illustrated in the example of the stress tensor, where we demonstrate the calculation for the $uu$ component explicitly. The other components can be calculated by following the same procedure. The value of $\braket{T_{uu}}$ is given as, 
  \be
  \braket{:T_{uu}:} = \braket{\p_u \phi(u) \p_u \phi(u)}
  \ee
  Now, according to the first step, we first separate the two points by a small amount, i.e, 
  \be
  \braket{:T_{uu}:} = \lim_{u' \to u}\braket{\p_u \phi(u') \p_u \phi(u')}
  \ee
  Now, one can pull out the derivatives and use, 
  \be
  \braket{:T_{uu}:} = \lim_{u' \to u} \p_u \p_{u'} \braket{\phi(u) \phi(u')}~.
  \ee
 Note that in the above, the quantity $\braket{\phi(u) \phi(u')}$ is nothing but the Green's function expressed in that particular coordinates.  
 Therefore, depending on the value of the Green's function, we can compute the expectation value of the Stress-Tensor explicitly. Remember that in the present situation, the expectation value will be calculated in the Minkowski state from the non-inertial observer. This identical procedure is exploited in obtaining the standard results, like Rindler observer in Minkowski spacetime or static observer in black hole spacetime (see, for example \cite{Birrell}).

\vskip 3mm 
 \noindent
 {\bf Thermal-Rindler:}
 In this case, the Minkowski state is thermal state with inverse temperature $\beta$ and the observer is the Rindler one. So we need to calculate $ \braket{:T_{\mc{U}\mc{U}}(\mc{U}):}_{\b} = \lim_{\mc{U}' \to \mc{U}} \p_{\mc{U}} \p_{\mc{U}'} \braket{\phi(\mc{U}) \phi(\mc{U}')}_{\b}$. Now since we are dealing with massless scalers, the Green's function is determined by expressing the relevant Minkowski counterpart in the Rindler coordinates. This has been done in (\ref{Wight-TR}). 
 
 Using (\ref{Wight-TR}) and upon carrying this computation we have,
  \be
  \braket{:T_{\mc{U}\mc{U}}(\mc{U}):}_{\b} = \lim_{\bar{\mc{U}} \to 0}\Big[\f{a^2}{48 \pi} + \f{\pi}{12 \b^2} e^{- 2 a \mc{U}} - \f{1}{4\pi \bar{\mc{U}}^2}\Big]~.
  \ee
  Here we denoted $\bar{{\mc{U}}}={\mc{U}}-{\mc{U}}'$.
Finally, extracting the finite part, we get first part of (\ref{Tuu-TR}). 
   
\vskip 2mm   
\noindent   
{\bf Rindler-Rindler:}
In this case the Minkowski state is vacuum state and the observer is Rindler-Rindler one.
   Proceeding in an exact similar manner, we would need to evaluate, 
   \be
    \braket{:T_{uu}(u):} = \lim_{u' \to u} \p_u \p_{u'} \braket{\phi(u) \phi(u')}~.
    \ee
The relevant Green's function is given by (\ref{Wight-RR}). Substitution of this in the above yields,
    \be
    \braket{:T_{uu}(u):} = \lim_{\bar{u} \to 0} \Big[\f{a^2}{48 \pi} + \f{a_1^2}{48 \pi} e^{- 2 a_2 u} - \f{1}{4\pi \bar{u}^2}\Big]~,
    \ee
with $\bar{u}=u-u'$.  We again see that the finite part of gives us the first part of stress tensor (\ref{TuuTvv}). 
   
In these derivations we must keep in mind that the way the point splitting technique has been chosen, it is not covariant. We must actually displace the second operator by an infinitesimal amount along the tangent on the manifold, which is equivalent to writing it as a parallel transport as explained in \cite{Birrell}. Doing this leads to an additional contribution of terms like $\q_{uu}$, etc. in eq.(\ref{Tuu-TR}) and eq.(\ref{TuuTvv}).  This has been done in the main results of the renormalised stress-tensor.
  
\section{\label{App6}Number Operator and all that}
One can relate the creation and annihilation operators of the Minkowski, Rindler and Rindler-Rindler using the Bogoliubov transformations, 
\begin{eqnarray}
c_q &= \intsinf \,d{p}  \Big(\a_{(21)pq} b_p + \b_{(21)pq}^* b_p^{\dagg}\Big)~;
\nonumber
\\     
b_q &= \intsinf \,d{p} \Big( \a_{(10)pq} a_p + \b_{(10)pq}^* a_p^{\dagg}\Big)~,
\label{Bogo-1}
\end{eqnarray}
where, $a (a^\dagg$), $b (b^\dagg$) and $c (c^\dagg$) are the annihilation (creation) operators corresponding to Minkowski, Rindler and Rindler-Rindler, respectively. The values for $\a$ and $\b$, for positive frequencies are given by (\ref{alpha}). 
The subscripts $(21)$ and $(10)$ in eq.(\ref{Bogo-1}) indicate that we are dealing with transformations between Rindler ($1$) $\to$ Rindler-Rinder ($2$) or Minkowski ($0$) $\to$ Rindler ($1$), respectively. The Bogoliubov transformation relating the Minkowski and Rindler-Rindler modes are, 
\begin{eqnarray}
\a_{(20)pq} &= \intsinf \,d{k} \Bigl[ \a_{(10)pk} \a_{(21)kq} + \b_{(10)pk} \b_{(21)kq}^* \Bigr]~;
\nonumber
\\
 \b_{(20)pq} &= \intsinf \,d{k} \Bigl[ \a_{(10)pk} \b_{(21)kq} + \b_{(10)pk} \a_{(21)kq}^* \Bigr].
\end{eqnarray}

Using the Bogoliubov transformations, as illustrated in Appendix \ref{App2}, we find the following values. 
\vskip 1mm
\noindent
 $\bullet$ \underline{$\braket{c_q c_r}_M$}: 
 \begin{equation}
  \braket{c_q c_r}_M = \intsinf\,d{p}\l[ \f{\a_{(21)pq} \b_{(21)pr}^* }{e^{2\pi p/a_1} - 1} +  \f{\a_{(21)pr} \b_{(21)pq}^* }{1 - e^{-2\pi p/a_1} }\r]~,
\end{equation}
where, 
\begin{eqn}
 &\intsinf\,d{p} \f{\a_{(21)pq} \b_{(21)pr}^* }{e^{2\pi p/a_1} - 1} \\
 &= \intsinf \f{\,d{p}}{e^{2\pi p/a_1} - 1} \l( -\f{1}{4\pi^2 a_2^2} \r) \f{\sqrt{rq}}{p} e^{\f{\pi}{2a_2}(q - r)} \\
 &\qquad\qquad \qquad\qquad\times\l( \f{a_2}{p} \r)^{\f{i (q + r)}{a_2}} \G\l( \f{i q}{a_2} \r) \G\l( \f{i r}{a_2} \r)  \\
 &= \l( - \f{\sqrt{rq}}{4\pi^2 a_2^2} \r) e^{\f{\pi}{2 a_2}(q - r)}  \l( \f{a_1 a_2}{2\pi} \r)^{\f{i(q + r)}{a_2}} \G\l( \f{i q}{a_2} \r) \G\l( \f{i r}{a_2} \r) \\
 &\qquad \qquad \qquad \qquad\times\G\l( -\f{i(q + r)}{a_2} \r) \z\l(- \f{i (q + r)}{a_2} \r)~,
\end{eqn}
and,
 \begin{eqn}
 &\intsinf \,d{p} \f{\a_{(21)pr} \b_{(21)pq}^*}{1- e^{- 2\pi p/a_1}}
 \\
 &=\l( -\f{\sqrt{rq}}{4\pi^2 a_2^2} \r)  e^{\f{\pi}{2a_2}(r -q)} \l( \f{a_1a_2}{2\pi } \r)^{\f{i (q+r)}{a_2}} \Gamma\l( \f{i q}{a_2} \r) \Gamma\l( \f{i r}{a_2} \r)  \\
 &\qquad \qquad \qquad \times\G\l( -\f{i (q + r)}{a_2}\r) \z\l( - \f{i (q + r)}{a_2} \r)~. 
 \end{eqn}
 
\noindent 
$\bullet$ \underline{$\braket{c_q^{\dagg} c_r}_M$}:
 \begin{eqn}
  \braket{c_q^{\dagg} c_r}_{\b} = \intsinf \,d{p} \biggl[ \f{\a_{(21)pq}^* \a_{(21)pr}}{e^{2\pi p/a_1} - 1} + \f{\b_{(21)pq} \b_{(21)pr}^*}{1 - e^{-2\pi p/a_1}}  \biggr]~, 
 \end{eqn}
 where, 
 \begin{eqn}
  &\intsinf \,d{p}  \f{\a_{(21)pq}^* \a_{(21)pr}}{e^{2\pi p/a_1} - 1} \\
  &=\f{\sqrt{rq}}{4\pi^2 a_2^2} e^{\pi (q + r)/2a_2} \G\l( \f{i r}{a_2} \r) \G\l( -\f{i q}{a_2} \r) \\
  &\qquad\qquad\times\l( \f{a_1a_2}{2\pi} \r)^{i (r - q)/a_2}\G\l( \f{r - q}{a_2} \r) \z\l( \f{r - q}{a_2} \r)~,
 \end{eqn}
  and, 
  \begin{eqn}
   &\intsinf \,d{p}  \f{\b_{(21)pq} \b^*_{(21)pr}}{1 - e^{-2\pi p/a_1}} \\
   &= \f{\sqrt{rq}}{4\pi^2 a_2^2} e^{-\pi (q + r)/2a_2} \G\l( \f{i r}{a_2} \r) \G\l( -\f{i q}{a_2} \r)  \\
   &\qquad \qquad \times \l( \f{a_1 a_2}{2\pi} \r)^{i (r - q)/a_2}\G\l( \f{r - q}{a_2} \r) \z\l( \f{r - q}{a_2} \r)~.
  \end{eqn}

 \noindent 
 $\bullet$ \underline{$\braket{c_{q} c^{\dagg}_{r}}_M$}: \\ \\
  This is directly evaluated from the commutation relation, 
 \begin{eqn}
  [c_q, c_r^{\dagg}] = c_q c_r^{\dagg} - c_r^{\dagg} c_q = \d_{qr}~.
 \end{eqn}

\noindent 
$\bullet$ \underline{$\braket{c_{q}^{\dagg} c^{\dagg}_{r}}_M$}:
  \begin{eqn}
  \braket{c_{q}^{\dagg} c^{\dagg}_{r}}_M
  = \intsinf \,d{p} \Biggl[  \f{\a^*_{(21)pq} \b_{(21)pr}}{e^{\f{2\pi p}{a_1}} - 1} + \f{\a_{(21)pq} \b^*_{(21)pr}}{1 - e^{-\f{2\pi p}{a_1}} }\Biggr]~,
  \end{eqn}
  where,
  \begin{eqn}
   &\intsinf\,d{p}  \f{\a^*_{(21)pq} \b_{(21)pr}}{e^{{2\pi p}/{a_1}} - 1} \\
   &=-\f{\sqrt{qr}}{4\pi^2 a_2^2}  \l( \f{a_1 a_2}{2\pi} \r)^{-\f{i (q + r)}{a_2}} \G\l( -\f{i q}{a_2} \r) \G\l( -\f{i r}{a_2} \r) \\
   &\qquad\qquad\times e^{\f{\pi}{2a_2}(q - r)}\G\l( -\f{i (q+r)}{a_2} \r) \z\l( -\f{i (q+r)}{a_2} \r)~,
  \end{eqn}
  and, 
  \begin{eqn}
   &\intsinf\,d{p}  \f{\a_{(21)pq} \b^*_{(21)pr}}{1 - e^{-{2\pi p}/{a_1}}} \\
   &= -\f{\sqrt{qr}}{4\pi^2 a_2^2} e^{-\f{\pi}{2a_2}(q - r)} \l( \f{a_1 a_2}{2\pi} \r)^{\f{i (q + r)}{a_2}} \G\l( \f{i q}{a_2} \r)  \\
   &\qquad \qquad \times \G\l( \f{i r}{a_2} \r) \G\l( \f{i (q+r)}{a_2} \r) \z\l( \f{i (q+r)}{a_2} \r)~.
  \end{eqn}

\section{Rotating frame in cylindrical coordinates}{\label{Cy}}
Previously we evaluated the Green's function in the rotating frame by substituting the trajectory of the detector in the Minkowski Green's function in Cartesian coordinates. In this appendix, we shall demonstrate the calculation of the same object, by expanding the field in the cylindrical coordinate system, i.e, using (\ref{cyl mode}). This does not give any new information, but here we shall be able to give the analytical expression of the response function in a much more convenient way as the integration can be done. So for completeness this will be discussed here which may be useful in some situations.

In rotating frame proper time adopted to cylindrical coordinates, the trajectory of the detector is given by \cite{Gutti:2010nv}
\be
\tilde{x}(\t) = \big(\g \t, \ \s, \ \g \O \t,\ 0\big)~.
\ee
Substituting of this in (\ref{new1-1}) yields
\begin{eqn}
G_{\b}^+(\breve{u}) &= \f{1}{4\pi^2} \sum_{m = -\infty}^{\infty} \intsinf\,d{q} \intinf\,d{k_z}\\
&\quad\times\f{q}{2\o} J_m^2(q \s) e^{i m \g \O \breve{u}}
 \Biggl[  \f{e^{-i \g \o \breve{u}}}{1 - e^{-\b \o}} + \f{e^{i \g \o \breve{u}}}{e^{\b \o} - 1}\Biggr]~.
\end{eqn}
Corresponding rate of the transition probability of the detector is then
\begin{eqn}
  R(E)=\int_{-\infty}^{\infty}d\breve{u}~e^{-iE\breve{u}}~G_\b^+(\breve{u})\equiv I_1+I_2 ~; 
 \end{eqn}
where, 
  \begin{eqn}
  I_1=&\frac{1}{(2{\pi})^2}\sum_{m=-{\infty}}^{\infty}\int_{-\infty}^{\infty}d\breve{u}\int_{0}^{\infty}\,d{q}\int_{-\infty}^{\infty}\,d{k_z}\\
  &\qquad\times \Biggl[\frac{q}{2{\o}}
  J_m^2(q{\s})\frac{\exp{[-i{\g}{\o}\breve{u}-{iE\breve{u}}+{im{\g}{\O}\breve{u}}}]}{1-e^{-{\b}{\o}}} \Biggr]~,
  \label{I1}
 \end{eqn}
 and 
\begin{eqn}
  I_2=\frac{1}{(2{\pi})^2}&\sum_{m=-{\infty}}^{\infty}\int_{-\infty}^{\infty}d\breve{u}\int_{0}^{\infty}\,d{q}\int_{-\infty}^{\infty}\,d{k_z} \\
  &\times\frac{q}{2{\o}}J_m^2(q{\s})\frac{\exp{[i{\g}{\o}\breve{u}-{iE\breve{u}}+{im{\g}{\O}\breve{u}}}]}{e^{{\b}{\o}}-1}~.
  \label{I2}
\end{eqn} 
These integrations are evaluated in the following way.

Let us start with (\ref{I1}). Performing the integral upon $\breve{u}$ first we get,
\begin{eqn}
I_1=&\frac{1}{2{\pi}}\sum_{m=-{\infty}}^{\infty}\int_{0}^{\infty}\,d{q}\int_{-\infty}^{\infty}\,d{k_z} \\
&\qquad\times \Biggl[\frac{q}{2{\o}}
J_m^2(q{\s})\frac{\d{[E+{\g\o}-{m{\g}{\O}}}]}{1-e^{-{\b}{\o}}} \Biggr]~.
 \end{eqn}
 From the value delta function and the relation $\o^2=q^2+k_z^2$ can write,
 \begin{align}
  &{\o}=(m-{\bar E})\O=\d;\qquad{\bar E}=E/(\g\O)\\
  &k_0={\pm}(\d^2-q^2)^{1/2}
 \end{align}
 where $k_0$ are the roots of $k_z$. As $E>0$ and $\o>0$, the value of $\d$ will be positive provided $m>{\bar E}$. Keeping this in mind the above is expressed as
 \begin{eqn}
 I_1=&\frac{1}{2{\pi}}\sum_{m={\bar E}}^{\infty}\int_{0}^{\infty}\,d{q}\int_{-\infty}^{\infty}\,d{k_z} \\
 &\times\frac{q}{2{\o}({1-e^{-{\b}{\o}}})}J_m^2(q{\s})\Big[\frac{\d({k_z}-{k_0})}{\g|{d\o}/{dk_z}|_{k_0}}\Big]
\end{eqn}
The two roots of $k_z$ give equal contribution, and as $k_z$ is real, the upper limit of $q$ can be set to $\d$ to evaluate the integral.
\begin{eqn}
 I_1&=\frac{1}{2{\pi}\g}\sum_{m={\bar E}}^{\infty}\int_{q=0}^{\d}\frac{q~dq}{({1-e^{-{\b}{\d}}})}\frac{J_m^2(q{\s})}{{({\d}^2-q^2})^{1/2}}\\ 
  &=\frac{1}{2 {\pi} {\g}}\sum_{m={\bar E}}^{\infty}\l(\frac{\d~({\s} {\d})^{2m}}{{\Gamma}(2m+2)}\r) \l(\frac{1}{1-e^{-{\b}{\d}}}\r) \\
  &\quad\times{}_1F_2{\Big[(m+1/2);(m+3/2),(2m+1);-(\s \d)^2\Big]}\\
\end{eqn}
\\
The second integral $I_2$, also corresponds to positive frequency as is given by, 
\begin{eqn}
  I_2=\frac{1}{(2{\pi})^2}&\sum_{m=-{\infty}}^{\infty}\int_{-\infty}^{\infty}d\breve{u}\int_{0}^{\infty}\,d{q}\int_{-\infty}^{\infty}\,d{k_z} \\
  &\times\frac{q}{2{\o}}J_m^2(q{\s})\frac{\exp{[i{\g}{\o}\breve{u}-{iE\breve{u}}+{im{\g}{\O}\breve{u}}}]}{e^{{\b}{\o}}-1}
\end{eqn}
Similarly we can write in terms of delta function,
\begin{eqn}
 I_2=\frac{1}{2{\pi}}&\sum_{m=-{\infty}}^{\infty}\int_{0}^{\infty}\,d{q}\int_{-\infty}^{\infty}\,d{k_z} \\
 &\times\frac{q}{2{\o}}J_m^2(q{\s})\frac{\d{(-{\g}{\o}+{E}-{m{\g}{\O}}})}{e^{{\b}{\o}}-1}
\end{eqn}
and,
\begin{align}
 &{\o}=(-m+{\bar E})\O=\d';\qquad{\bar E}=E/(\g\O)\\
  &k_0={\pm}(\d'^2-q^2)^{1/2}
\end{align}
As $\o>0$ the value of $\d'$ will be positive when $m<{\bar E}$. So, the sum will be
\begin{eqn}
 I_2=\frac{1}{2{\pi}}&\sum_{m=-{\infty}}^{\bar E}\int_{0}^{\infty}\,d{q}\int_{-\infty}^{\infty}\,d{k_z} \\
 &\times\frac{q}{2{\o}({e^{{\b}{\o}}}-1)}J_m^2(q{\s})\l[\frac{{\d}({k_z}-{k_0})}{\g|{d\o}/{dk_z}|_{k_0}}\r]
\end{eqn}

If we change $m\rightarrow-m$ the Bessel function $J_{-m}^2(q\s)=J_{m}^2(q\s)$ remains the same, as $m$ is an integer. Using this fact, $I_2$ can be written as,
\begin{eqn}
I_2 &=\frac{1}{2{\pi}\g}\sum_{m={-\bar E}}^{\infty}\int_{0}^{\d'}\,d{q} \frac{q}{({e^{{\b}{\d'}}}-1)}\frac{J_m^2(q{\s})}{{({\d'}^2-q^2})^{1/2}}\\ 
 &=\frac{1}{2 {\pi} {\g}}\sum_{m=-{\bar E}}^{\infty}\l(\frac{\d' ~({\s} {\d'})^{2m}}{{\Gamma}(2m+2)}\r) \l(\frac{1}{e^{{\b}{\d'}}-1}\r)\\
 &\qquad \times {}_1F_2{\Big[(m+1/2);(m+3/2),(2m+1);-(\s \d')^2\Big]}~.
\end{eqn}

Finally, the response function can be obtained by evaluating the sums in $I_1$ and $I_2$. 
One can check numerically that the sum over $m$ converges.

 \end{appendix}

\end{document}